\documentclass{jfm}
\pdfoutput=1
\let\today\relax

\makeatletter
\def\ps@pprintTitle{%
    \let\@oddhead\@empty
    \let\@evenhead\@empty
    \def\@oddfoot{\footnotesize\itshape
         {\hfill  June 2023}  \today}%
    \let\@evenfoot\@oddfoot
    }
\usepackage[T1]{fontenc}
\usepackage[round]{natbib}
\usepackage{graphicx}

\usepackage{amsmath}
\usepackage[outdir=./]{epstopdf}
\usepackage{bm}
\usepackage{tikz}
\usepackage{xcolor}

\usepackage{floatrow}
\usepackage{adjustbox}
\usepackage{bm}
\usepackage{makecell}
\usepackage{graphicx}
\usepackage{amssymb}
\usepackage[section]{placeins} 
\usepackage{stackengine}
\usepackage{scalerel}
\usepackage{subfigure}
\usepackage{lastpage}
\usepackage{comment}

\newcommand{\Realset}{\mathbb{R}}
\newcommand{\blackline}{\tikz{\draw[black, line width=1.5pt] (0,0) -- (0.5cm,0);}}

\newcommand{\bluedots}{\tikz{\draw[blue, line width=2pt,dotted] (0,0) -- (0.5cm,0);}}
\definecolor{mygray}{RGB}{128,128,128}
 \newcommand{\greydots}{\tikz{\draw[mygray, line width=1.5pt,dotted] (0,0) -- (0.5cm,0);}}

\newcommand{\reddashed}{\tikz{\draw[red, line width=1.5pt,dashed] (0,0) -- (0.5cm,0);}}

\definecolor{mygreen}{RGB}{0, 128, 0} 
\newcommand{\greenbar}{%
  \tikz{
    \draw[mygreen, line width=0.5pt] (-0.1cm,-0.1cm) -- (0.1cm,0.1cm);
    \draw[mygreen, line width=0.5pt] (0.1cm,-0.1cm) -- (-0.1cm,0.1cm);
    \draw[mygreen, line width=1.5pt] (-0.25cm,0cm) -- (0.25cm,0cm);
  }%
}
\definecolor{reddark}{rgb}{0.71,0.14,0.07}
\newcommand\xrowht[2][0]{\addstackgap[.5\dimexpr#2\relax]{\vphantom{#1}}}
\stackMath
\newcommand\reallywidehat[1]{%
\savestack{\tmpbox}{\stretchto{%
  \scaleto{%
    \scalerel*[\widthof{\ensuremath{#1}}]{\kern-.6pt\bigwedge\kern-.6pt}%
    {\rule[-\textheight/2]{1ex}{\textheight}}
  }{\textheight}%
}{0.5ex}}%
\stackon[1pt]{#1}{\tmpbox}%
}

\newcommand{\dprime}{\prime \prime}

\newcommand{\cf}{{C_f}}
\newcommand{\cfa}[1]{C_{f#1}}

\includecomment{jpdf}
\excludecomment{jpdf}

\newcommand{\transpose}{\intercal}
\newcommand*\diff{\mathop{}\!\mathrm{d}}

\renewcommand{\Re}{\Rey}
\newcommand{\Nu}{Nu}
\newcommand{\Pe}{\Pen}
\renewcommand{\Pr}{\Pran}
\newcommand{\St}{St}

\title{Breaking the Reynolds Analogy: Decoupling Turbulent Heat and Momentum Transport via Spanwise Wall Oscillation in Wall-Bounded Flow}

\author{Lou Guérin \aff{1}  \corresp{\email{lou.guerin@univ-poitiers.fr}}, Cédric Flageul\aff{1}, Laurent Cordier\aff{1}, Stéphane Grieu\aff{2}, Lionel Agostini\aff{1}}

\affiliation{\aff{1}Institut Pprime, CNRS, Université de Poitiers, ISAE-ENSMA, 11 Bd. Marie et Pierre Curie, Site du Futuroscope, TSA 41123, 86073 Poitiers Cedex 9, France
\aff{2}Laboratoire PROMES (UPR CNRS 8521), Université de Perpignan Via Domitia, rambla de la thermodynamique, 66100 Perpignan, France}

\begin{document}
\maketitle

 \begin{abstract}

 This work investigates spanwise wall oscillation (SWO) as a method to preferentially enhance heat transfer over drag in turbulent channel flow. Direct numerical simulations at $\Rey_\tau=180$ and $\Pr=1$ show set of wall-oscillation parameters reducing drag also decrease heat transfer similarly, maintaining coupled transport. However, large period ($T^+=500$) and amplitude ($W^+=30$) induce substantially greater heat transfer intensification, increasing 15\% versus only 7.7\% drag rise. This Reynolds analogy breaking enables preferential elevation of heat transport over momentum. FIK identity analysis reveals negligible impact of forcing terms on dissimilarity. Instead, differences arise from the solenoidal velocity and linear temperature equations. Both the turbulent shear stress and heat flux are amplified near the wall under oscillation. However, the heat flux intensifies more substantially, especially at its peak. This preferential enhancement of the near-wall heat flux, exceeding the shear stress amplification, facilitates greater thermal transport augmentation relative to the friction increase. Results demonstrate  that spanwise wall oscillation can preferentially intensify heat transfer beyond drag, providing a promising technique for improving heat exchanger. Further work should optimize the period and amplitude of the oscillation and elucidate the underlying physics of this dissimilar heat transfer control.

\end{abstract}

\section{Introduction}
\label{seq:intro}

Turbulent flows dictate the performance characteristics of numerous industrial equipment and environmental applications. One important consequence of turbulence is to increase the mixing momentum leading to high friction drag on surfaces. The increase relative to laminar conditions easily reaches factors of 10-100, depending on the Reynolds number of the flow. In many applications, the friction drag is extremely influential to the operational effectiveness of the device or process \citep{2}. This applies especially to transport, involving either self-propelling bodies moving in a fluid or fluids being transported in ducts and pipes \citep{7}. There is significant pressure to reduce transport-related emissions, of which friction drag is a major constituent \citep{5}. On the other hand, enhancing the turbulent fluxes within the wall-bounded region is generally beneficial for heat transfer \citep{6}. Thus, in the case of heat exchangers, a balance needs to be found between drag-induced losses and heat transfer. For a wide variety of engineering systems involving cooling or heating processes, enhancing heat exchanger performance represents a pivotal technological challenge for greater efficiency, consistent with industrial and societal needs for cost-effective energy transfer.

For many years, controlling the boundary layer to decrease drag has been an active research area. One promising technique is imposing spanwise oscillation on the wall \citep{drag4,20,21,22}, as reviewed extensively by \citet{2}.  
Both simulations and experiments show that a spanwise oscillation can substantially reduce drag, around 40-50\% at low Reynolds number \citep{15,leschziner2012}. The oscillating wall motion introduces a time-varying spanwise strain near the wall. This disrupts streak formation and breakdown, weakens the quasi-streamwise vortices, and thickens the viscous sublayer \citep{drag4,18,drag1}. This alters the near-wall turbulence, reducing momentum mixing and thus shear stress at the wall. The drag reduction depends on SWO parameters such as amplitude, frequency, and waveform. While extensive research has aimed to determine optimal oscillation parameters for maximal turbulent drag reduction, the accompanying effects on heat transfer have received comparatively less focus. Oscillations increasing drag may also strengthen heat transfer due to the connection between momentum and heat transport. There is great interest in decoupling these mechanisms to substantially increase heat transfer while keeping drag low. Studying the heat transfer response to oscillatory wall forcing via simulations is therefore needed. This will clarify if similar control parameters exist for drag rise and heat transfer increase. Such knowledge will help develop oscillation-based strategies that simultaneously intensify heat transfer while restricting frictional penalties.

This study investigates the dynamics of dissimilar variations in heat transfer and drag in a turbulent channel flow with imposed spanwise wall oscillation. Direct numerical simulations (DNS) demonstrate SWO control inducing dissimilarity, the first instance of such results shown with DNS accuracy. To further examine the mechanisms involved, systematic decomposition of the friction coefficient and Nusselt number is performed. This enables detailed examination of the complex dynamics governing the response to oscillatory forcing. Obtaining drag reduction within a channel flow has been the focus of several investigations. It was shown that SWO is a reliable control strategy to reduce drag over large surfaces \citep{drag1,drag2,drag3,drag4,Marusic2021}. Several studies have focused on identifying the optimal set of parameters for maximizing the drag reduction \citep{Gatti2013}. 

Let $u_\tau$ and $\nu$ be the friction velocity and kinematic viscosity, respectively, it was demonstrated that one of the best parameter set for reducing drag is $T^+=100$ and $W^+=12$, where $T^+=\frac{T u_\tau^2}{\nu}$ and $W^+=\frac{W}{u_\tau}$ are the dimensionless period and amplitude of the oscillations, respectively. When control is applied, there is a transient phase of 2-3 oscillation periods in length, leading to the state of lowest drag. Once the low drag state is reached, phase variations synchronized with the wall oscillation period can be observed. The drag undergoes a strengthening and a weakening phase twice over each actuation cycle. The effect of the oscillation has been shown to decrease with increasing friction Reynolds number ($\Re_\tau= \frac{H u_\tau}{\nu}$) based on the half height $H$ of the channel \citep{leschziner2012,Marusic2021,Gatti2016,Hurst2014}.\\

Previous studies \citep{2,drag4,Marusic2021} have determined that the optimal oscillation period for maximum drag reduction is around $T^+\approx 100$. At this value, the largest reductions in skin friction are achieved through interactions between the oscillatory wall motion and near-wall turbulent structures. For example, results from direct numerical simulations  at Reynolds number $\Re_\tau=200$ demonstrate a maximum drag reduction of 44.7\% using $T^+\approx 100$ and amplitude $W^+=27$ \citep{drag4}. The optimal period appears to be relatively robust across different Reynolds numbers, with similar values found in turbulent channel flows up to $\Re_\tau=6000$ \citep{Marusic2021}. However, the amplitude required for maximum drag reduction increases with Reynolds number. It is notable that within the range of oscillation parameters ($0 \leq W^+ \leq 30$, $0 \leq T^+ \leq 300$) investigated by \citet{drag4} at $\Re_\tau=200$, drag increase was not observed. To the best of the authors' knowledge, the only documented instance, where an overall increase in drag at a low Reynolds number is evident, is in simulations conducted by \citet{drag6}, specifically at an extended period of $T^+ = 500$, with $W=0.8 Q_x$ where $Q_x$ is the fixed flow rate in the streamwise direction, and a corresponding $\Re_\tau$ of 200. Notably, this higher value of period revealed significantly more pronounced variations in the periodic equilibrium drag compared to shorter periods. In contrast, the study by \citet{Marusic2021}, employing a similar period of $T^+\approx600$, showcased drag reduction that proved increasingly effective with rising $\Re_\tau$ within the range $951 \leq \Re_\tau \leq 12800$. Major differences between the studies by \citet{drag6} and \citet{Marusic2021} could explain their contrary results on the drag variations. First, the control strategies differ, as \citet{drag6} investigate sinusoidal oscillatory spanwise wall motion while \citet{Marusic2021} study forcing in the form of travelling waves. Also, the travelling waves utilize weaker amplitude for the actuation compared to the sinusoidal oscillations. Furthermore, \citet{drag6} performed simulations at low Reynolds numbers where drag is primarily driven by near-wall turbulence structures. However, at higher Reynolds numbers studied by \citet{Marusic2021}, large-scale structures develop in the outer layer which strengthen and increasingly contribute to drag \citep{agostini2017multi,agostini2021statistical}. \citet{Marusic2021} show drag reduction is obtained by weakening the effect of these outer scales, which are absent at low Reynolds numbers. Although at first glance the results of \citet{Marusic2021} seem to contradict observations by \citet{drag6}, this difference likely originates from both the type of forcing employed and the flow regime.

There have been limited attempts to control heat transfer using spanwise wall oscillation. \citet{Fang2009} performed LES at $\Re_\tau=180$ and $\Pr=0.72$ on a weakly compressible flow at $\text{Mach}=0.5$, using oscillation periods near the optimal value and varying the amplitude from $6.35 \leq W^+ \leq 19.05$. They showed averaged wall heat flux can be reduced with appropriate parameters but increases in most cases. Temperature and streamwise velocity streaks were found to be consistent despite drastic changes from oscillations. \citet{Fang2010} later proved momentum and heat transport consistency, highly correlated with turbulent motions. Using in-phase oscillations \citep{Fang2010_active}, significant drag increase occurred, mainly in the transient stage. For all forms tested, heat transfer variations were highly similar to drag. \citet{Ni2016} extended these results to higher Mach number ($\text{Mach}=2.9$). They introduced a  corrected version of the wall-heat flux, on which the Stokes solution is subtracted. This corrected wall-heat flux is also shown to vary in a similar fashion than the drag.

When considering a configuration with a passive scalar and mixed boundary conditions (MBC) for the temperature \citep{thermo1,thermo7}, one can expect heat transfer to behave similarly to the drag, at least for Prandtl numbers near unity where streamwise velocity and temperature are strongly correlated \citep{Fang2011}. If this hypothesis is valid, heat transfer increase may not occur within the oscillation parameter ranges tested by \citet{drag3}. However, results  from \citet{drag6} suggest increased heat transfer could be achieved at higher periods.

Different control strategies  have been proposed in order to achieve what is referred to as dissimilar heat transfer \citep{DHT1,DHT2,DHT3}. The goal is to achieve a heat transfer increase while reducing the drag, or at least while minimising its increase.  The comparison,  performed by \citet{thermo6}, of the different possible boundary conditions for the temperature in a turbulent channel flow suggest that this choice may have a significant impact on whether the control can easily achieve dissimilar heat transfer or not. In \citet{DHT4}, the authors  compared different types of temperature boundary conditions, namely Uniform Heat Generation (UHG), Constant Heat Flux (CHF) and Constant Temperature Difference (CTD) by using the Fukagata-Iwamoto-Kasagi (FIK) identities \citep{drag5} and its heat transfer extension \citep{DHT5}. From this investigation, it results that the best scenario to obtain dissimilarity is when CTD thermal boundary conditions are applied and the Prandtl number is far from unity. 

\citet{DHT1} demonstrated a highly effective drag reduction and heat-transfer enhancement strategy using wall blowing and suction control in turbulent channel flow. At friction Reynolds numbers of $\Re_\tau=100,\ 150, \ \text{and} \ 300$ with a passive scalar temperature field and unity Prandtl number, they achieved up to $24 \%$ drag reduction along with over $50 \%$ increase in heat transfer. This was attained even with the unfavorable unity Prandtl number and UHG thermal boundary conditions, which typically make dissimilarity between drag and heat transfer difficult. However, it should be noted that while wall blowing/suction has shown promising results, its practical implementation is far more challenging compared to wall motion based strategies.

\citet{DHT2} employed a streamwise travelling wave-like wall deformation strategy at $\Re_\tau=180$. The temperature was modeled as a passive scalar with CTD thermal boundary conditions and $\Pr=1$. To quantify the dissimilarity achieved, they introduced the analogy factor $A_n=\frac{\St/\St^0}{\cfa{}/\cfa{}^0}$, where $\St$ and $\St^0$ are the actuated and unactuated heat transfer coefficients and $\cfa{}$ and $\cfa{}^0$ the actuated and unactuated skin friction coefficients, respectively. Their goal was obtaining $A_n > 1$ as large as possible while ensuring heat transfer enhancement ($\frac{\St}{\St^0} > 1$). Using optimized parameters, results from \citet{DHT2} attained a time averaged analogy factor of $1.13$ with this control strategy.

The present study aims to investigate the effect of spanwise wall oscillation towards dissimilar heat transfer.
For that, Direct Numerical Simulations of a passive scalar temperature with Mixed Boundary Conditions are performed at $\Re_\tau=180$ and $\Pr=1$.
After a description of the formulation and numerical conditions in section \ref{seq:config}, validation will be carried out in section \ref{seq:validation} by comparing the obtained DNS statistics with references at $\Re_\tau=180$ and observing the differences between each of the studied meshes. This will be followed by an introduction and validation of the control strategy in section \ref{seq:SWO}.
The results will then be presented in section \ref{seq:Results} where
a first step will be to find parameters outside the range studied in \citet{drag4}, allowing a heat transfer increase. The parameters $T^+ = 500$ and $W^+ = 30$ will serve as the initial focus of investigation, given that this period has been demonstrated to cause an overall increase in drag according to findings in \citet{drag6}.
 The FIK identity decompositions \citep{drag5} will then be investigated in order to identify the different components contributing to heat transfer and drag variations.
The main purposes of this work remain the validation of the configuration and the extension of spanwise wall oscillation to dissimilar heat transfer control.

\section{Flow conditions and simulation details}
\label{seq:config}
Results presented herein arise from a DNS for a canonical channel flow at $\Re_{\tau}\approx 180$. The flow domain is defined as :

$\Omega = \left\{  \boldsymbol{x}=\begin{pmatrix}
x \ y \ z\end{pmatrix}^\transpose \in \Realset^3 \ | \ x \in [0,L_x], \ y \in [0,L_y], \ z \in [0,L_z]  \right\}$   with $\boldsymbol{x}=\begin{pmatrix}
x \ y \ z\end{pmatrix}^\transpose$ representing the streamwise, wall normal and spanwise components, respectively (see figure \ref{fig:schematic}). The velocity, pressure and temperature fields are given by $\bm{u}(\bm{x}, t)=\begin{pmatrix}
u(\bm{x}, t) \ v(\bm{x}, t) \ w(\bm{x}, t) \end{pmatrix}^\transpose,\  p(\bm{x}, t)$ and $T(\bm{x}, t)$. 

The evolution of the velocity $\bm{u}$ is given by the incompressible Navier-Stokes equations, and that of the temperature $T$ corresponds to the passive transport of a scalar by the velocity $\bm{u}$. Assuming a flow density equal to 1, the following set of governing equations are obtained:

\begin{equation}
\begin{aligned}
& \left\{
\begin{aligned}
&\nabla \bm{\cdot} \bm{u} = 0 \\
&\frac{\partial \bm{u}}{\partial t} + \left( \bm{u} \bm{\cdot} \nabla \right) \bm{u} = -\nabla p + \nu \Delta \bm{u} \\
&\frac{\partial T}{\partial t} + \bm{u} \bm{\cdot} \nabla T = \alpha \Delta T,
\end{aligned}
\right. \\
& \label{eq:NS}
\end{aligned}
\end{equation}
where 
$\nu$ is the kinematic viscosity and 
$\alpha$ is the thermal conductivity.
Hereafter, bulk quantities are defined as  $\displaystyle \chi_b = \dfrac{1}{|\Omega |}\int_\Omega \langle\chi\rangle_t  \ \diff\Omega$ for any field $\chi$, where $|\Omega|$ is the fluid inner volume of the channel and $\langle\cdot\rangle_t$ the time average. By extension, $\langle \cdot \rangle_{x,z,t}$ denotes the average over time and the directions $x$ and $z$.
The flow rate $U_b$ is therefore obtained for $\chi=u$ and the bulk temperature $T_b$ for $\chi=T$.

In numerical simulations, periodic boundary conditions are imposed in the streamwise and spanwise directions of the flow, as well as homogeneous Dirichlet boundary conditions on the walls for both velocity and temperature (see below).
By definition, $\chi_\text{w}$ represents the boundary conditions imposed on any variable $\chi$ on the walls, i.e. $\chi_{\text{w}} = \left\{\chi(\bm{x})\mid x \in [0,L_x], \ z \in [0,L_z]\ \text{and}\ y=0 \ \text{or} \  y=2H \right\}$ where $H$ is the half height of the channel.

Concerning the thermal boundary condition, several options are available. Here, a Mixed Boundary Condition (MBC) is imposed, see \citet{thermo1,thermo7,thermo5} for a complete presentation. When MBC boundary conditions are specified, an averaged constant uniform heat flux $q_w$ is applied on the walls (see figure \ref{fig:schematic}) while assuming that the temperature fluctuations at the wall are null.
Therefore, the average temperature at the walls $T_\text{w}$ is independent of time and, due to the global heat balance for constant heat flux, it increases linearly in the streamwise direction ($ \langle T_{\text{w}} \rangle_t = Ax + T_0$). For this type of thermal boundary condition, \citet{thermo7} and \citet{thermo5} demonstrated that an appropriate non-dimensional form of the temperature is defined as: 

\begin{equation}
\Theta=\frac{T- \langle T_{\text{w}} \rangle_t}{T_b -\langle T_{\text{w}} \rangle_t }.
\label{eq:deftheta}
\end{equation}

It can be shown that the scaling temperature $\langle T_{\mathrm{w}} \rangle_t-T_b$ is constant by applying Newton's law of cooling,
$q_w=h(\langle T_{\mathrm{w}} \rangle_t-T_b)$ where the heat transfer coefficient $h$ is constant, and using the constant flux assumption.
Furthermore, the assumption of zero fluctuations at the wall can be directly incorporated in the prescription of the boundary condition:

\begin{equation}
\Theta_\text{w} = 
\Theta(T=T_\text{w}) = 
\frac{T_\text{w} - \langle T_\text{w} \rangle_t}{T_b -\langle T_\text{w} \rangle_t } = 
\frac{T_\text{w}^\prime}{T_b -\langle T_\text{w}\rangle_t} = 0.
\end{equation}
Finally, the bulk temperature is constant ($\Theta_b=\Theta(T=T_b)=1$) ensuring the thermal stationary condition.

\subsection{Non-dimensionalisation}
The dimensionless Navier-Stokes equations are defined as:

\begin{equation}
\begin{aligned}
& \left\{
\begin{aligned}
& \nabla^\star \bm{\cdot} \bm{u}^\star = 0 \\
&\frac{\partial \bm{u}^\star}{\partial t^\star} + \left( \bm{u}^\star \bm{\cdot} \nabla^\star \right) \bm{u}^\star = -\nabla^\star p^\star + \frac{1}{\Re} \Delta^\star \bm{u}^\star + \bm{f}^\star \\
&\frac{\partial \Theta}{\partial t^\star} + \bm{u}^\star\bm{\cdot} \nabla^\star \Theta + A^\star u^\star = \frac{1}{\Pe} \Delta \Theta
\end{aligned}
\right. \\
& \label{eq:NS_adim}
\end{aligned}
\end{equation}
with 
$$\Theta=\frac{T- \langle T_{\text{w}} \rangle_t}{T_b -\langle T_{\text{w}} \rangle_t }  ,\ \boldsymbol{x^\star}=\frac{\boldsymbol{x}}{H}, \  t^\star= t \frac{U_b}{H}, \ p^\star=\frac{p}{U_b^2}, \  \boldsymbol{u^\star}=\frac{\boldsymbol{u}}{U_b}, \ A^\star = A \frac{H}{T_b -\langle T_{\text{w}} \rangle_t}. $$

Introducing the variable $\Theta$ given by equation \eqref{eq:deftheta} into the temperature equation generates a forcing term  $f_\Theta=Au$.
The approach to implement this forcing term follows the constant bulk temperature (CBT) method, wherein $\Theta_b=1$ is mandated at each iteration.
For velocity, a constant flow rate (CFR) strategy is imposed, fixing $U_b=\frac{2}{3}$ at each iteration. The corresponding forcing term $\bm{f^\star}  = \begin{pmatrix}f^\star_x & 0 & 0\end{pmatrix}^\transpose$ arises from the mean pressure gradient driving the flow. This compensates for viscous friction to achieve steady state. The CFR procedure is common and will not be detailed here, as it is thoroughly described in \citet{thermo7}.

Subsequently, for simplicity, notations without superscript 
$\cdot^\star$
will refer to dimensionless values. Notations with superscript 
$\cdot^+$
will denote values scaled by wall units, i.e. non-dimensionalised by $u_{\tau}, \ \Theta_{\tau} = \frac{\alpha}{u_\tau}
\frac{\partial \langle \Theta_\text{w} \rangle_{x,z,t}}{\partial y}$ and $\Re_{\tau}$, the friction velocity, friction temperature and friction Reynolds number, respectively.  

The two key dimensionless numbers characterizing the flow are the Reynolds number $\Re=\frac{U_b H}{\nu}$ and the Péclet number $\Pe=\Re \times\Pr$, where $\Pr$ is the Prandtl number defined as $\Pr=\frac{\nu}{\alpha}$. 
For this study, the Prandtl number is set to $\Pr=1$ to obtain higher similarity between the velocity and temperature equations as $\Pe=\Re$.
In this case, identical boundary conditions are utilized for both velocity and temperature. This enables simpler analysis, as differences in the variations of drag and heat transfer induced by actuation will not originate from disparities in the unactuated boundary conditions or Prandtl number effects.
This channel flow configuration was selected for its simplicity and ability to induce substantial similarity between the streamwise velocity and temperature fields. 

In this configuration, the sole dissimilarities between the streamwise velocity and temperature fields that could potentially facilitate attaining dissimilar heat transfer are:

\begin{itemize}
\item[--] The source term difference, $f_x$ is uniform in space while $f_\Theta$ is not.

\item[--] The divergence-free condition of the continuity equation applies to velocity but not temperature.

\item[--] The linearity of the convective term in the temperature equation does not apply to streamwise velocity.
\end{itemize}

The friction coefficient, $\cfa{}$, and Nusselt number, $\Nu$, characterize drag and heat transfer, respectively. Computation of both dimensionless parameters utilizes the forcing terms, $f_x$ and $f_\Theta$. Despite employing a distinct definition of $\Theta_b$, calculation intricacies are comprehensively explained in \citet{thermo7}. For validation, current simulations are compared to reference uncontrolled cases. Correspondence between simulations and canonical cases provides a measure of implemented model and numerical method veracity.

To enable forthcoming validations, the definition of the time-wise and phase-wise averages must be clearly established. Any general field variable $\chi$, representing velocity components, temperature or pressure can be decomposed as

\begin{equation}
 \chi(x,y,z,t) = \overline{\chi} (y) + \chi^{\prime}(x,y,z,t), \end{equation}
where $\overline{\chi}=\langle \chi \rangle_{x,z,t}$ is the time and space average at a particular wall-normal location $y$, and $\chi^{\prime}$ are the stochastic fluctuations in absense of wall oscillations.

For the actuated field, the stochastic fluctuations are noted as $\chi^{\dprime}$ and are obtained by removing the phase average of the raw field:

\begin{equation}
\chi^{\prime \prime}(x,y,z,t)=\chi(x,y,z,t)-\widetilde{\chi}(\epsilon,y),
\label{eq:phase_stat}
\end{equation}
with the phase average $\widetilde{\chi}(\epsilon,y)$ defined as:

\begin{equation}
\widetilde{\chi}(\epsilon,y)=\frac{1}{N} \sum_{n=1}^N \langle \chi(x,y,z,\epsilon+nT) \rangle_{x,z},
\label{eq:phase_stat_2}
\end{equation} 
and where $\epsilon \in \{0,..,T\}$ is the phase and $N$ is the number of cycles over which the averaging is performed.

Finally, another notation is introduced to define the periodic fluctuations:

\begin{equation} 
\widehat{\chi}(\epsilon,y)=\widetilde{\chi}(\epsilon,y)-\overline{\chi}(y).
\label{eq:phase_stat_3}
\end{equation}

\subsection{Numerical simulation details}
The open source in-house Xcompact3d framework \citep{Xcompact1,Xcompact2,Xcompact3} is used to perform numerical simulations. Sixth-order compact finite difference schemes are used for spatial discretisation and a third-order explicit Runge-Kutta scheme is chosen for time integration. The condition of zero velocity divergence is ensured using a fractional step method, where a Poisson equation for the pressure gradient is solved with 3D FFTs  \citep{Xcompact1}. For validation, two mesh resolutions, namely $S_1$ and $S_2$, are investigated and compared  with reference simulations (see table \ref{tab:simus}).

\begin{table}
    \scalebox{0.7}{%
    \begin{tabular}{|c||c|c|c|c|c|c|c|c|c|}\hline
    Cases &$(Lx,Ly,Lz)$  
    &$\Delta x^+$& $\Delta y^+_{min} $& $\Delta y^+_{max} $ &$\Delta z^+ $ & $\Delta t^+$ & stat collection period $(t^+)$ & $\Re_\tau$ \\
    \hline

    \textbf{\citet{thermo8}} & $(12.8,2,6.4)$ &  $3.0$&  $0.20$& $5.93$& $3.0$ & NA & $3960$ &$180$ \\

      \textbf{\citet{thermo9}} & $(6.4,2,3.2)$ &  $1.1$&  $0.05$& $0.97$& $1.1$ & NA & $1677$ &$180$ \\
   
    \hline\xrowht[()]{5pt}
        $S_1$ &  $(8,2,4)$ & $4.96$& $0.29$& $4.13$& $3.53 $& $1.1 \ 10^{-2}$  &$5613$ &$178$ \\
  $S_2$ &$(24,2,6)$ & $10.68$ &$0.43$ &$6.16$ &$5.34$&  $3.41 \ 10^{-2}$ & $3790$ & $179$ \\
   
  \hline
        
    \end{tabular}} \\
    
    \caption{DNS unactuated computational conditions at $\Re_{\tau} \approx 180$.}
    \label{tab:simus}
\end{table}

The first reference simulation  \citep{thermo8} is utilized to validate the velocity statistics, while the second reference simulation  \citep{thermo9} verifies the temperature statistics.

Mesh $S_1$ employs a fine spatial discretization but on a limited domain size. In contrast, Mesh $S_2$ encompasses a larger domain for the periodic directions, although with coarser grid resolution compared to Mesh $S_1$. Mesh $S_1$ exhibits approximately 1.5 times higher resolution in the wall-normal and spanwise directions and 2.15 times higher resolution in the streamwise direction.

In \citet{thermo10}, in which the Xcompact3D code was also used, it was shown that an increase in numerical dissipation in the streamwise direction was, in some cases, equivalent to  increasing the spatial resolution. This increase of the numerical dissipation was carried out using spectral vanishing viscosity of fourth order accuracy \citep{LAMBALLAIS20113270}. Extra dissipation is therefore added in the streamwise direction on mesh $S_2$. It will be demonstrated in section \ref{seq:validation} that in such a configuration, $S_2$ yields results similar to the higher resolution grid of $S_1$.

\section{Numerical simulations validation}
\label{seq:validation}
\begin{figure}
\centering
\subfigure[]{\label{fig:umean}\includegraphics[trim= 0 0 0   0, clip=true,width=0.44\textwidth]{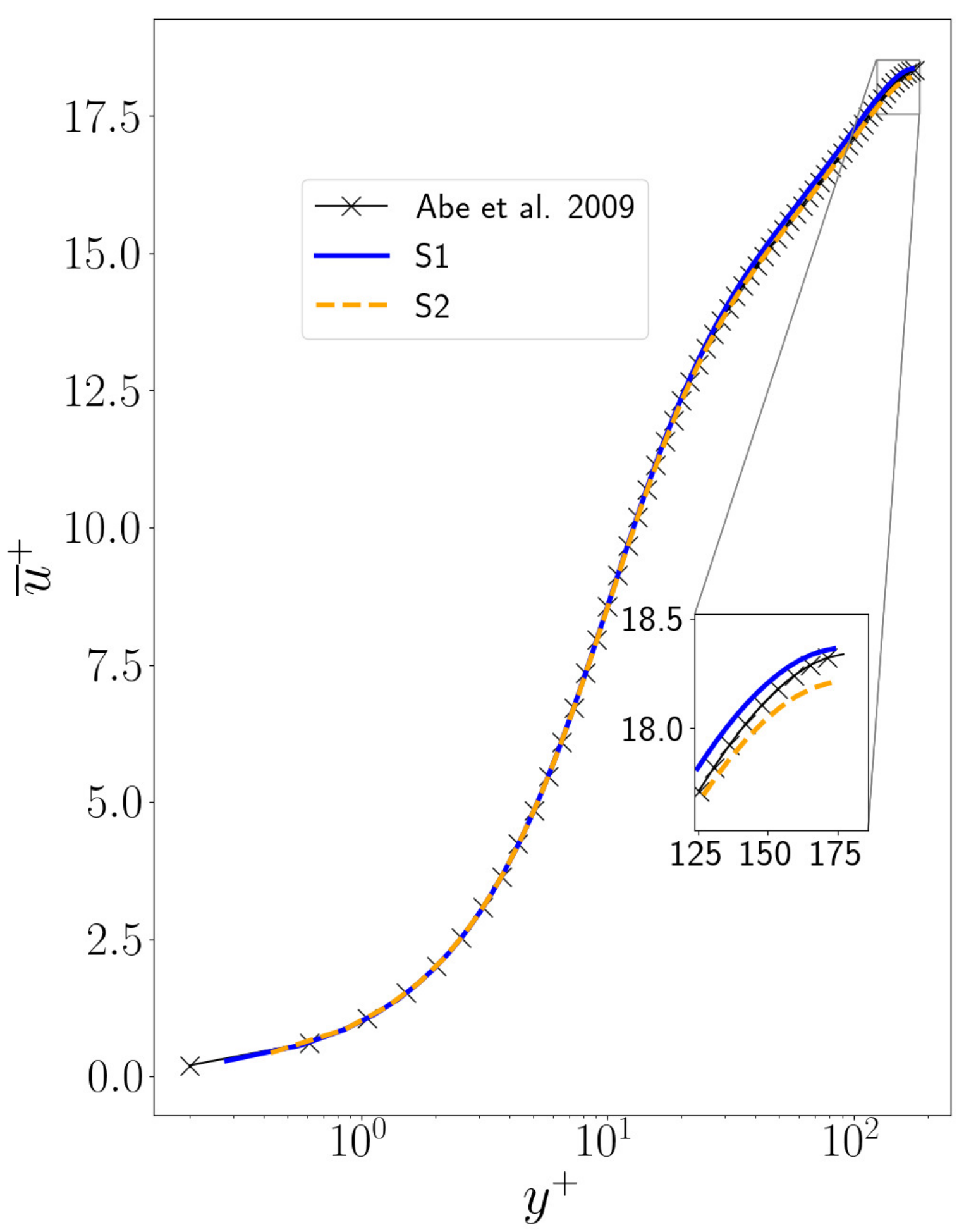}}
\subfigure[]{\label{fig:thetamean}\includegraphics[trim= 0 0 0   0, clip=true,width=0.44\textwidth]{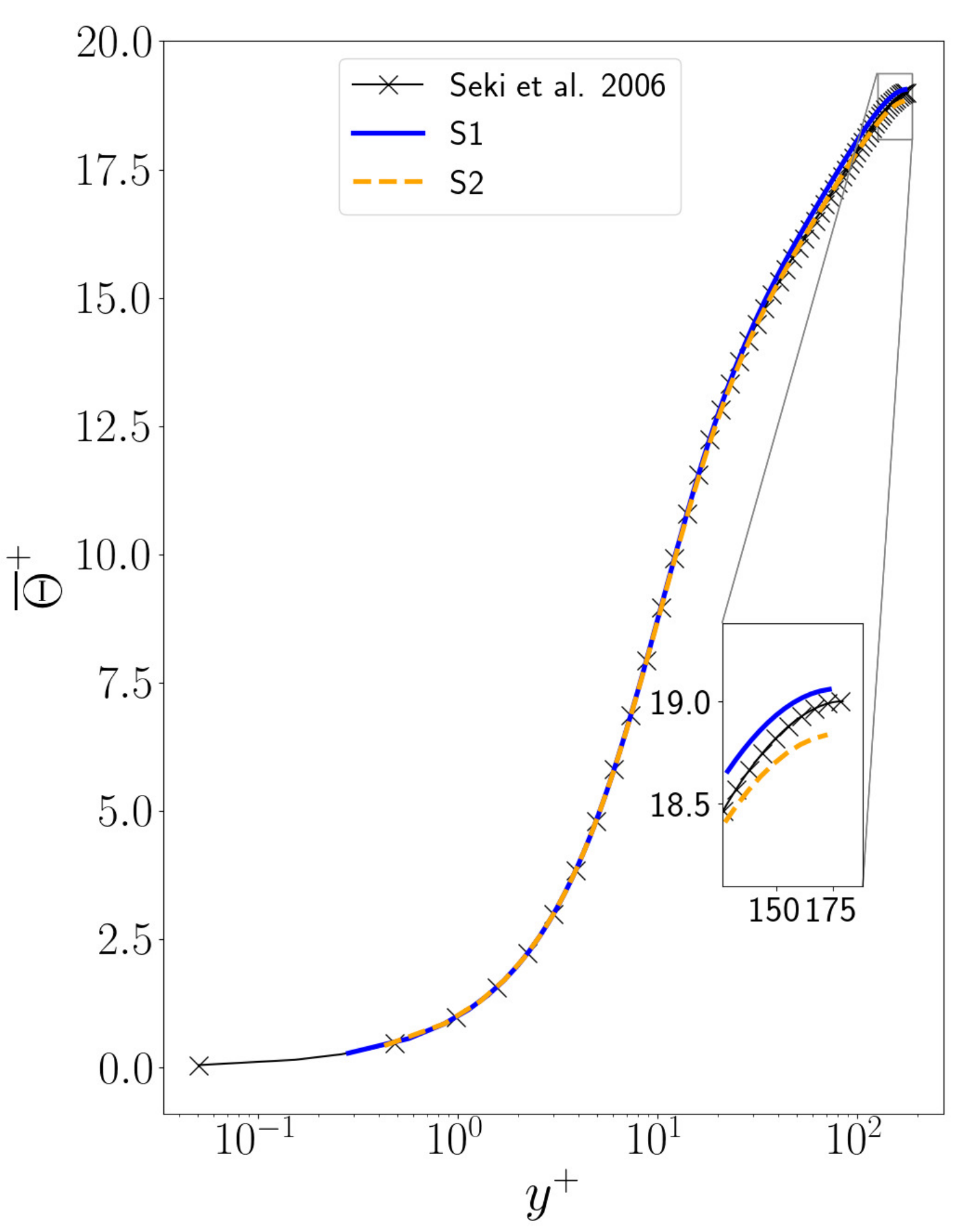}}
\caption{Wall-normal distribution of the mean: (a) streamwise velocity and (b) temperature. Statistics obtained from simulations on $S_1$ and $S_2$ are represented by plain blue line and dashed orange line, respectively. Reference results \citep{thermo8,thermo9} shown by the black line.}
  \label{fig:Mean}
\end{figure}

\begin{figure}
\centering
\subfigure[]{\label{fig:uuprime}\includegraphics[trim= 0 0 0 0, clip=true,width=0.44\textwidth]{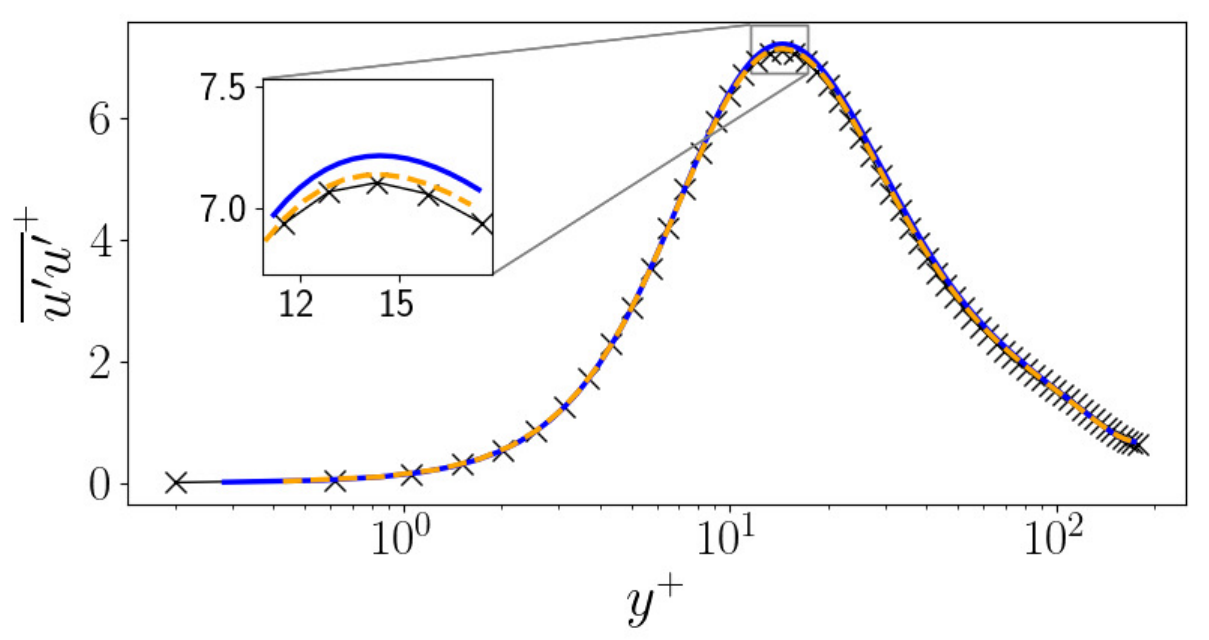}}
\subfigure[]{\label{fig:uvprime}\includegraphics[trim= 0 0 0  0, clip=true,width=0.44\textwidth]{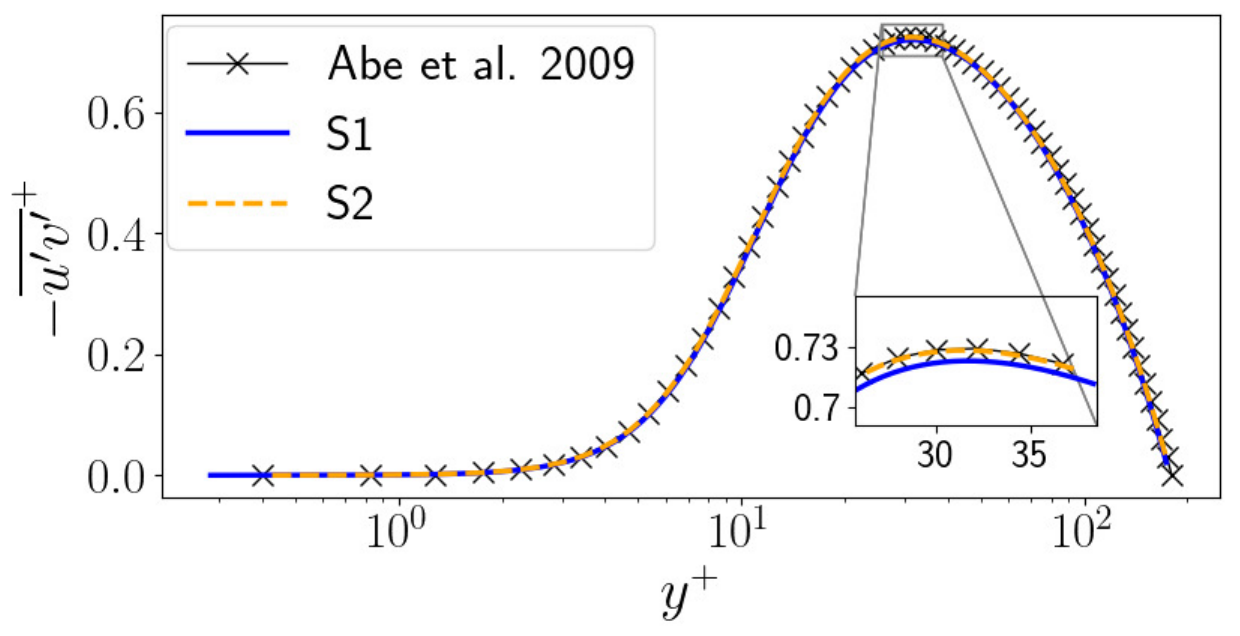}}
\subfigure[]{\label{fig:vvprime}\includegraphics[trim= 0 0 0   0, clip=true,width=0.44\textwidth]{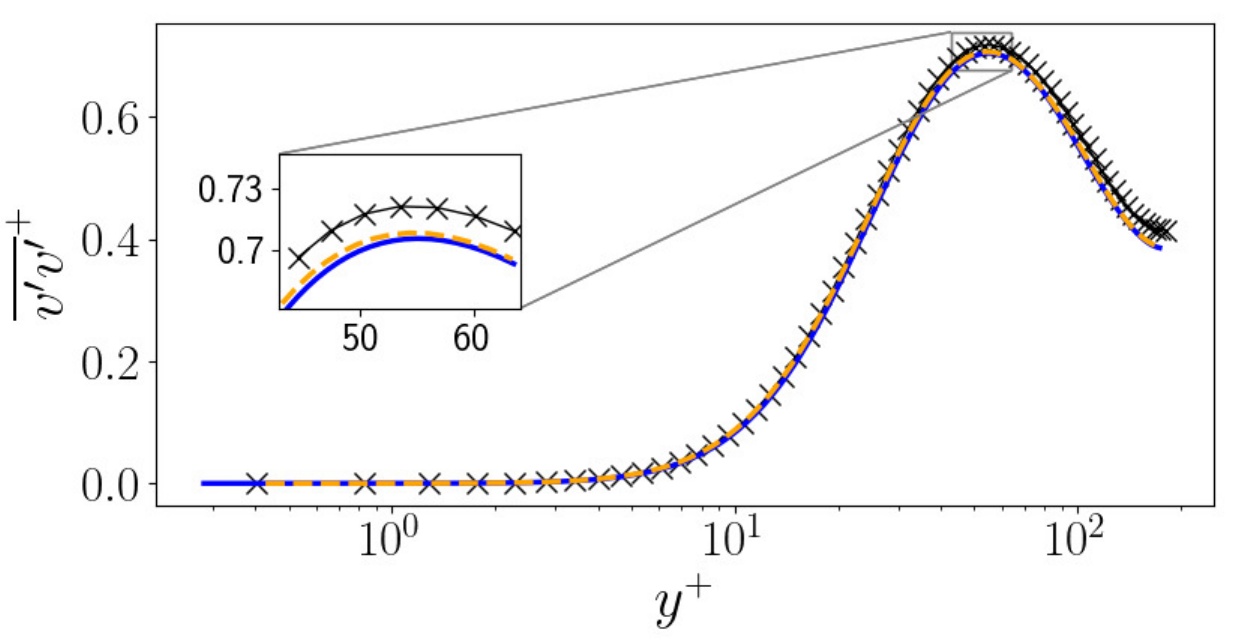}}
\subfigure[]{\label{fig:wwprime}\includegraphics[trim= 0 0 0   0, clip=true,width=0.44\textwidth]{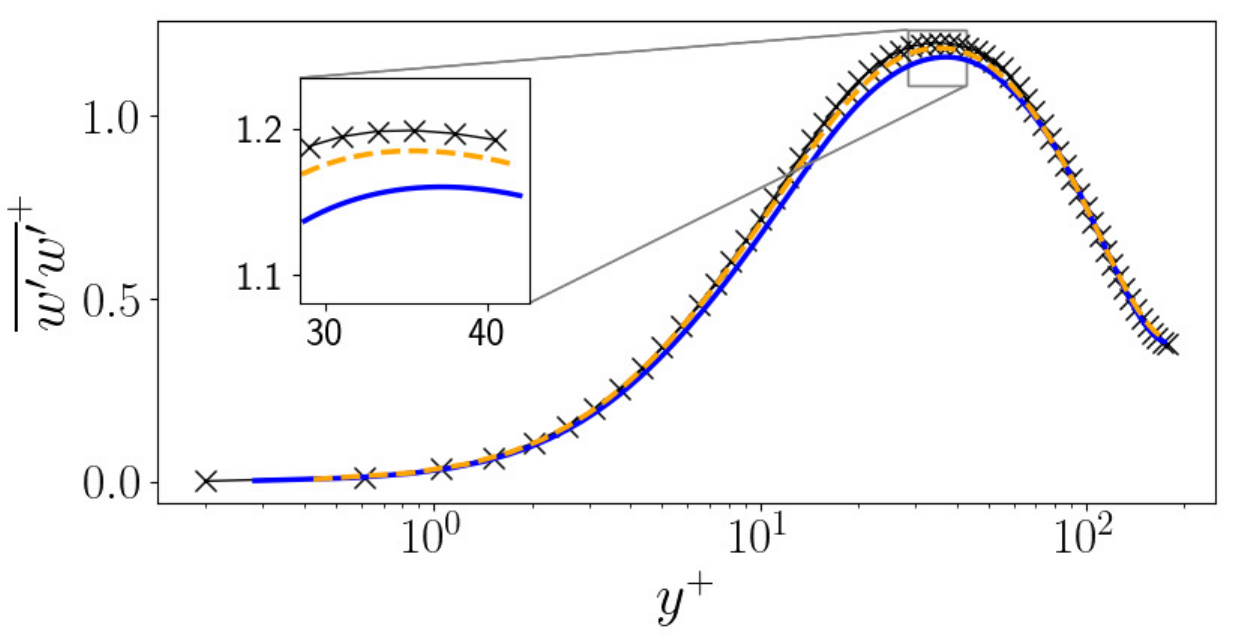}}
\caption{Wall-normal distribution of the Reynolds stress components: (a)  $\overline{u^{\prime}u^{\prime}}$, (b) $-\overline{u^{\prime}v^{\prime}}$, (c) $\overline{v^{\prime}v^{\prime}}$ and (d) $\overline{w^{\prime}w^{\prime}}$. 
See figure \ref{fig:Mean} caption for legend details.
}
  \label{fig:vel_stats}
\end{figure}

\begin{figure}
\centering
\subfigure[]{\label{fig:thetathetaprime}\includegraphics[trim= 0 0 0   0, clip=true,width=0.44\textwidth]{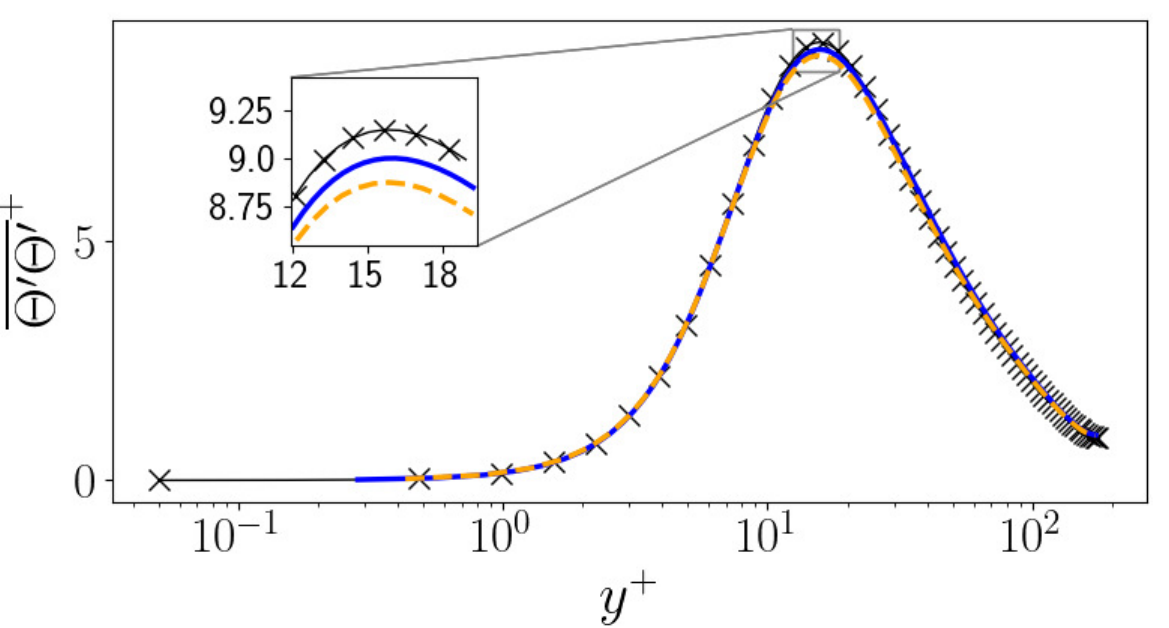}}\\
\subfigure[]{\label{fig:vthetaprime}\includegraphics[trim= 0 0 0   0, clip=true,width=0.44\textwidth]{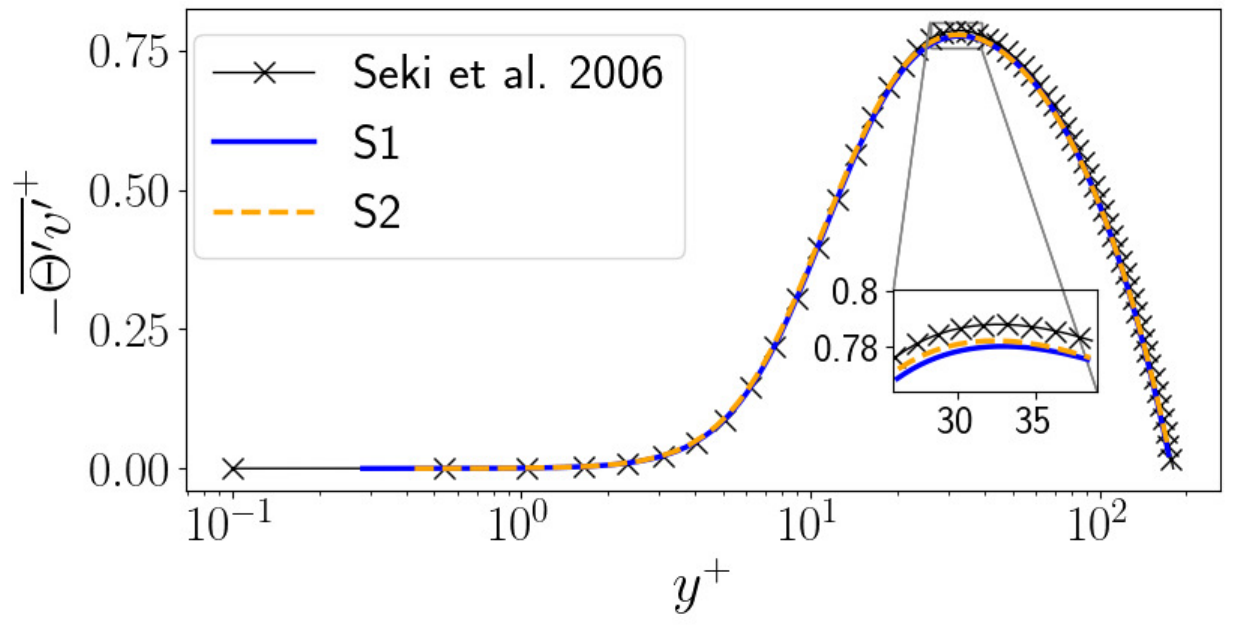}}
\subfigure[]{\label{fig:uthetaprime}\includegraphics[trim= 0 0 0   0, clip=true,width=0.44\textwidth]{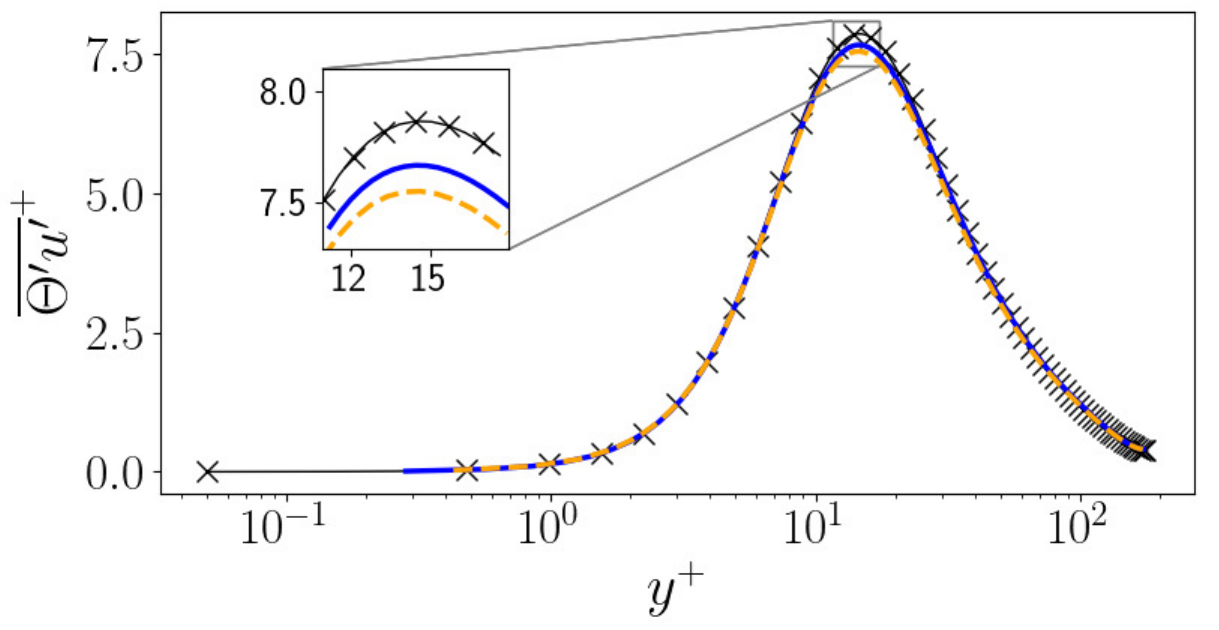}}

\caption{ Wall-normal distribution of the turbulent heat transport: (a)  $\overline{\Theta^{\prime}\Theta^{\prime}}$, (b) $-\overline{\Theta^{\prime}v^{\prime}}$ and (c) $\overline{\Theta^{\prime}u^{\prime}}$  at $\Pr=1$.  
See figure \ref{fig:Mean} caption for legend details.
}
  \label{fig:temp_stats}
\end{figure}

\begin{figure}
    \subfigure[]{ \label{fig:Cf}
    \includegraphics[trim= 2.5cm 1cm 0cm 0cm, clip=true,width=0.44\textwidth,height=0.5\textwidth]{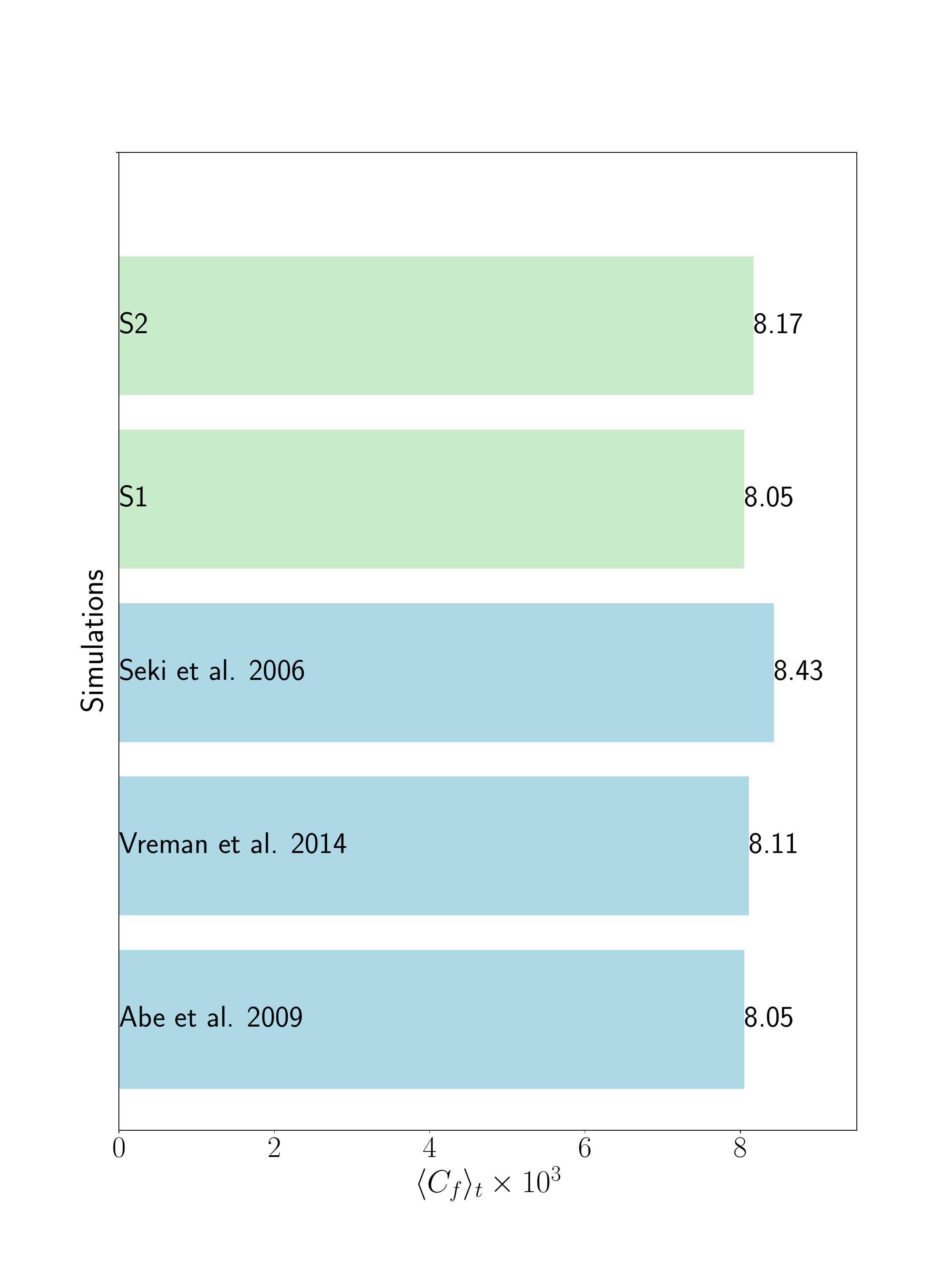}
  }
  \subfigure[]{ \label{fig:Nu}
    \includegraphics[trim= 2.5cm 1cm 0cm 0cm, clip=true,width=0.44\textwidth,height=0.5\textwidth]{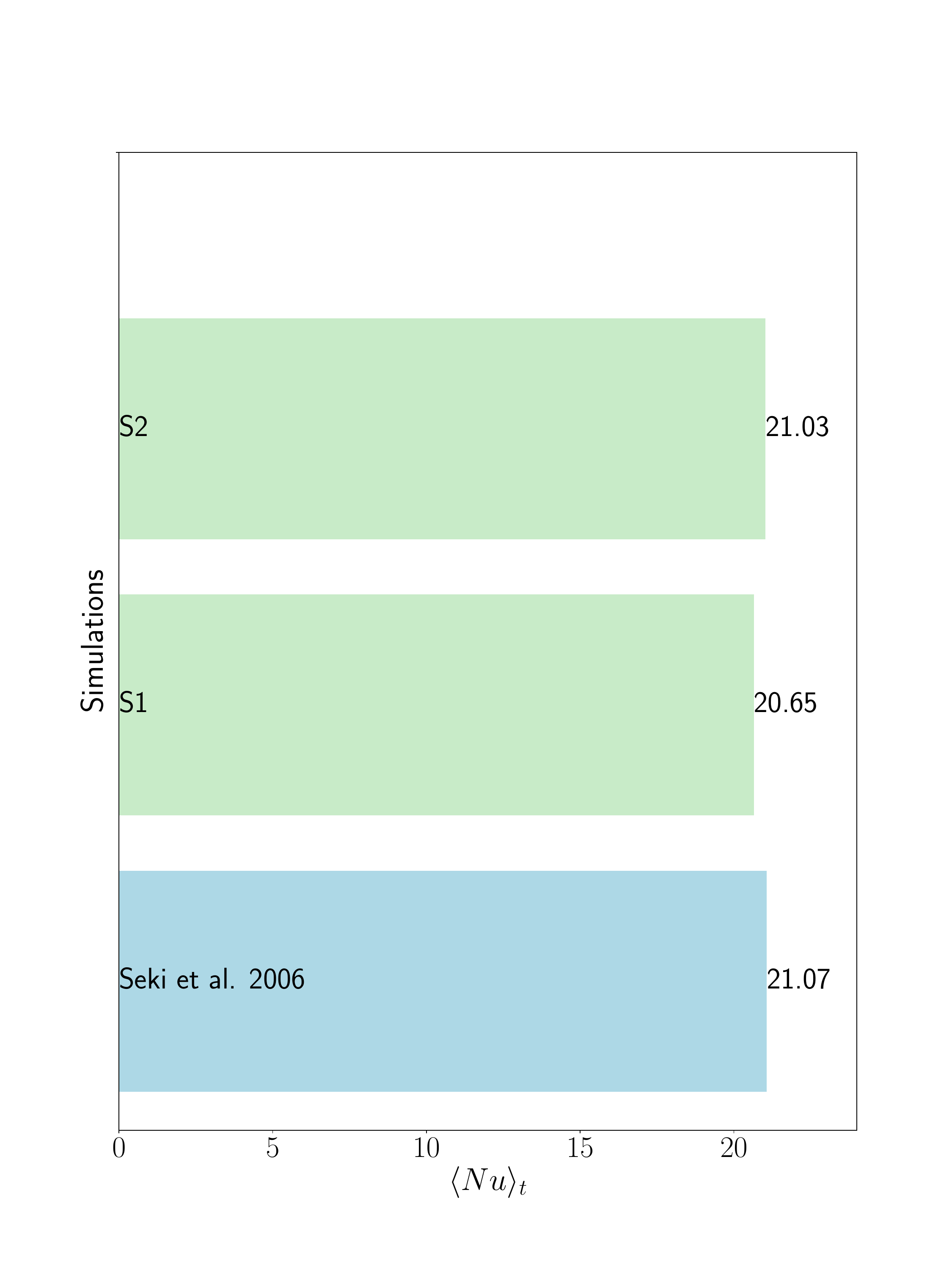}
    }

  \caption{(a) Mean friction coefficient $\langle \cf \rangle_t$  on $S_1$, $S_2$ (green bins) and on different reference data (blue bins), 
  (b) Mean Nusselt number $\langle Nu \rangle_t$  on $S_1$, $S_2$ (green bins) and on 
  reference data (blue bin). 
  }
  \label{fig:Nu_Cf}
\end{figure}

Figures \ref{fig:umean}, \subref{fig:thetamean} convey the wall-normal distributions of the mean streamwise velocity and temperature, respectively. Simulations utilizing the S1 mesh are denoted by the solid blue lines, while the orange dashed lines indicate simulations on the S2 mesh. The black solid lines with crosses represent the reference data of \citet{thermo8} providing the benchmark velocity profile and \citet{thermo9} supplying the temperature profile. For both variables, the current simulation results exhibit strong agreement with the reference data, with only negligible discrepancies in the channel center that are statistically inconsequential.

Figures \ref{fig:vel_stats} and \ref{fig:temp_stats} convey the second-order statistical moments for velocity and temperature, respectively. Figure \ref{fig:vel_stats} shows the Reynolds stresses while figure \ref{fig:temp_stats} displays the temperature variance and turbulent heat fluxes. For the Reynolds stresses, both simulations exhibit excellent agreement with the reference data, with minor deviations mainly for the $S_1$ discretization. Specifically, the $S_1$ mesh slightly overpredicts the peak of $\overline{u^\prime u^\prime}$ (figure \ref{fig:uuprime}) and underpredicts $\overline{w^\prime w^\prime}$ (figure \ref{fig:wwprime}). The profiles of the wall shear stress $\overline{u^\prime v^\prime}$ (figure \ref{fig:uvprime}) are nearly superimposed for both discretizations, indicating the accuracy of the prediction of this quantity.

Regarding  temperature statistics in figure \ref{fig:temp_stats}, unlike to the Reynolds stresses, the $S_1$ mesh results show marginally better agreement with the reference data compared to the $S_2$ mesh. However, both simulations slightly underestimate the peaks of $\overline{u^\prime \Theta^\prime}$ and $\overline{\Theta^\prime \Theta^\prime}$. The discrepancy with the simulations of \citet{thermo9} could result from their finer mesh resolution. However, their data were collected over a more limited spatial and temporal domain compared to the present studies. This smaller sampling could cause statistics that are not fully converged. Overall, figures \ref{fig:vel_stats} and \ref{fig:temp_stats} highlight robust agreement with the reference statistics \citep{thermo8,thermo9}. 
Slight differences are likely attributable to statistical convergence.

\begin{figure}[h!]
\centering
\subfigure[]{\label{fig:PSD_compare_x_u}\includegraphics[trim= 0 0 0  0, clip=true,width=0.32\textwidth]{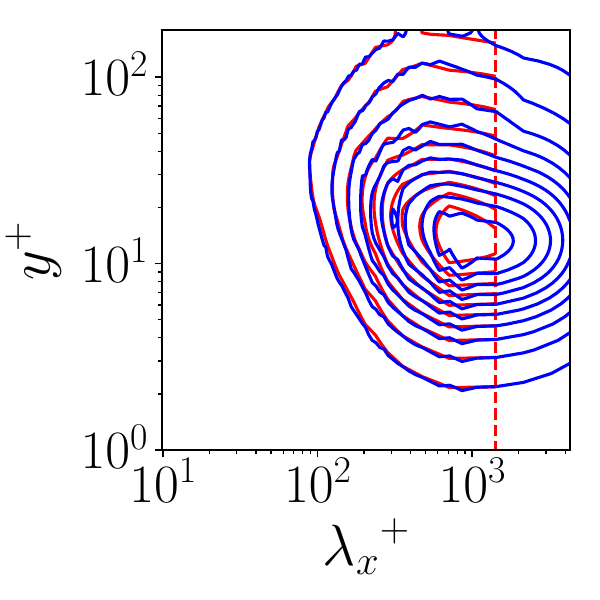}}
\subfigure[]{\label{fig:PSD_compare_z_u}\includegraphics[trim= 0 0 0  0, clip=true,width=0.32\textwidth]{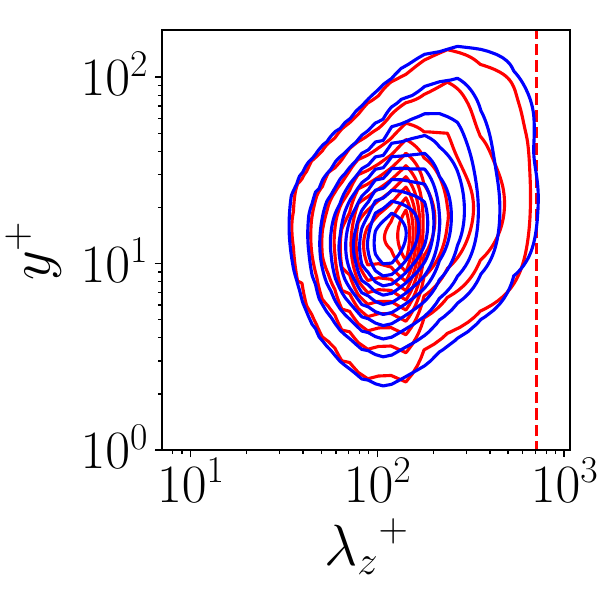}}
\subfigure[]{\label{fig:PSD_compare_int_u}\includegraphics[trim= 0 0 0  0, clip=true,width=0.32\textwidth]{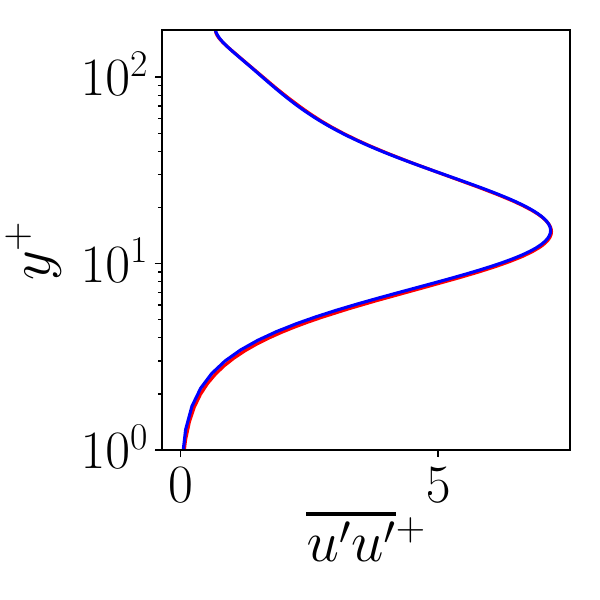}}

\subfigure[]{\label{fig:PSD_compare_x_phi}\includegraphics[trim= 0 0 0  0, clip=true,width=0.32\textwidth]{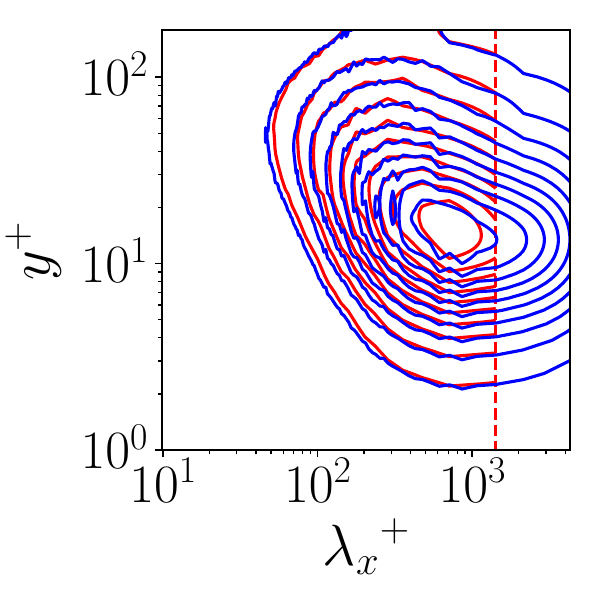}}
\subfigure[]{\label{fig:PSD_compare_z_phi}\includegraphics[trim= 0 0 0  0, clip=true,width=0.32\textwidth]{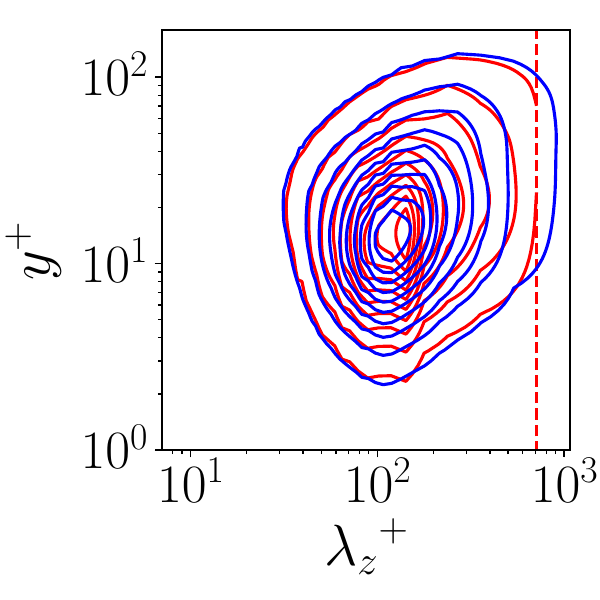}}
\subfigure[]{\label{fig:PSD_compare_int_phi}\includegraphics[trim= 0 0 0  0, clip=true,width=0.32\textwidth]{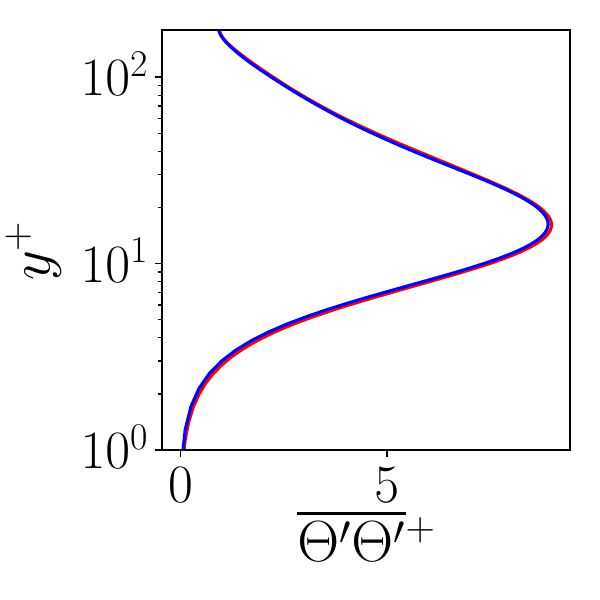}}

\caption{Pre-multiplied Power Spectral Densities of (a,b,c): the streamwise velocity field. $\left[ k_d \Phi_{uu}(\lambda_d,y)\right]^+$: (a) in the streamwise direction, $d=x$, (b) in the spanwise direction, $d=z$, and (c) $\overline{u^{\prime}u^{\prime}}^+$ wall-normal distribution. (d,e,f): same quantities for the temperature field. Red isolines are the results obtained on mesh $S_1$ and blue isolines, on mesh $S_2$. Isoline levels span linearly between the minimal and maximal values. Red vertical dashed lines show the limit in domain size of mesh $S_1$.}.
\label{fig:PSD_compare_u_phi}
\end{figure}

Figure \ref{fig:Nu_Cf} presents the temporal average of the friction coefficient $\langle \cfa{} \rangle_t$  and Nusselt number $\langle \Nu \rangle_t$, which characterize drag and heat transfer, respectively.  Comparison against three distinct sets of reference database serves to validate these quantities. Examining the $S_2$ mesh results, $\langle \cfa{} \rangle_t$ lies between the reference values, with discrepancies of 3.2\% with \citet{thermo9}, 1.4\% against \citet{thermo8}, and 0.9\% with \citet{DNS2}. The $\langle \cfa{} \rangle_t$ value from the simulation on the $S_1$ mesh  is equal to the one measured by \citet{thermo8}. Regarding the Nusselt number, the $S_2$ configuration exhibits a discrepancy of 0.2\% with the results of \citet{thermo9}, while the $S_1$ configuration shows a larger deviation of 2.0\%.
 Both discretizations accurately predict the relevant statistics in the present study, with the additional observation that the $S_2$ mesh demonstrates a lower error in computing the Nusselt number.

The pre-multiplied power spectral densities and the wall-normal distribution of the variance of the streamwise velocity and temperature are conveyed by figure \ref{fig:PSD_compare_u_phi}.
The left column shows spectra in the streamwise direction while the middle column shows spectra in the spanwise direction. The right column displays the wall-normal variation of $\overline{\chi^\prime \chi^\prime}$ obtained by integrating the pre-multiplied power spectral density $\Phi_{\chi^{\prime}\chi^{\prime}}$ along the wavelength: 

\begin{equation}
\label{eq:phi_uu}
\overline{\chi^{\prime} \chi^{ \prime}} = \int_0^{\infty} \Phi_{\chi^{\prime} \chi^{\prime}}\,\diff f = \int_0^{\infty} \lambda \Phi_{\chi^{\prime} \chi^{\prime}}\,\diff \log \lambda.
\end{equation}

These figures show the impact of the domain sizes of meshes $S_1$ and $S_2$. The isolines and second order stresses match closely between the two meshes, indicating increased resolution on $S_2$ would not necessarily improve the precision of results. This underpins the earlier hypothesis that the reference statistics from \citet{thermo9} may not be fully converged. As expected, velocity spectrum peaks occur at $\lambda_x^+ \approx 1000$ streamwise and $\lambda_z^+ \approx 100$ spanwise. The small $S_1$ domain size does not capture low-frequency content.

Analysis of the streamwise velocity and temperature spectra reveals that the streamwise domain length of mesh $S_2$ is marginally adequate to capture the relevant flow physics. Quantitative assessment of the second-order statistics shows mesh $S_2$ provides slightly improved precision for predicting the turbulence quantities relevant for this study. Based on these assessments, mesh $S_2$ is selected for the remainder of this work, examining the effects of spanwise wall oscillation on variations in drag and convective heat transfer. The marginally enhanced domain size and sufficient resolution of $S_2$ are expected to provide more accurate quantification of the oscillation-induced changes in wall shear stress and Nusselt number under the actuation conditions considered.

\section{Control strategy: Spanwise wall oscillation}
\label{seq:SWO}

\subsection{Numerical simulation}
\label{seq:config_actuation}
\begin{figure}
\center
\includegraphics[scale=0.5]{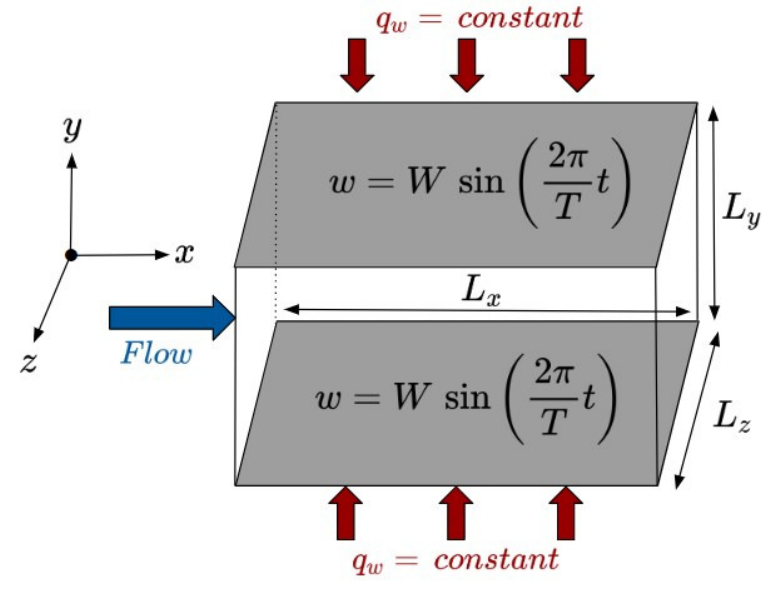}\\
\caption{Schematic of the flow configuration.}
\label{fig:schematic}
\end{figure}

The spanwise wall oscillation of period $T$ and velocity amplitude $W$ (see figure \ref{fig:schematic}) are achieved by specifying the following time-dependent boundary condition for the spanwise velocity at the wall position:

\begin{equation}
\label{eq:SWO}
w_{\text{w}}(t_n) = W\sin(\frac{2\pi}{T} t_n) \ \ \ \text{where} \ t_n= n \Delta t. \end{equation}

Here, $t_n$ denotes the discrete time at time step index $n$, and $\Delta t$ is the constant time step size. This formulation imposes a harmonic oscillation in the spanwise wall velocity.

To enable examination of the actuation effects, phase-averaged statistics are defined per equations \eqref{eq:phase_stat}, \eqref{eq:phase_stat_2} and \eqref{eq:phase_stat_3}. These leverage the inherent periodicity of the oscillatory forcing to delineate the statistical evolution across discrete phases within the actuation cycle. Such detailed interrogation is necessary to elucidate the fundamental physical phenomena governing the fluctuations in drag and heat transfer.

Due to the imposed oscillations and their periodic time-dependence, three variants of wall units may be employed for non-dimensionalization in the actuated flow:
\begin{itemize}
    \item[--] $\chi^+_{nom}$ where the normalization is based on the mean baseline friction velocity $\overline{u_{\tau,nom}}$.
    \item[--] $\chi^+_{ac}$ where the mean actual friction velocity $\overline{u_{\tau,ac}}$ is used for scaling.
    \item[--] $\chi^+_{ac,p}$ where scaling is based on the phase-averaged actual friction velocity $\widetilde{u_{\tau,ac}}$.
\end{itemize}
    
For both computational meshes, the Courant-Friedrichs-Lewy (CFL) number was constrained under $0.5$. To satisfy this CFL limit with Mesh $S_2$, the time step was reduced to compensate for CFL increase induced by the actuation, as given in table \ref{tab:simus_ac}. This table provides details on the actuated flow simulations.

\begin{table}
    \scalebox{0.7}{%
    \begin{tabular}{|c||c|c|c|c|c|}\hline
    Cases &  $\Delta t^+_{nom}$ & stat collection period $(t^+_{nom})$ & number of cycles & number of phases per cycle for snapshots / stats \\
   
    \hline\xrowht[()]{5pt}
        $S_1$ & $1.1 \ 10^{-2}$  &$11500$ &$23$ & $32$ / $288$ \\
        $S_2$&  $1.56 \ 10^{-2}$  &$11500$ &$23$ & $32$ / $400$\\
   
  \hline
        
    \end{tabular}} \\
    
    \caption{DNS actuated computational conditions, with oscillation parameters $T^+_{nom}=500$ and $W^+_{nom}=30$.}
    \label{tab:simus_ac}
\end{table}

Relative to the unactuated case, attaining statistical convergence necessitates extended integration intervals, as precise representations are critical for each phase of the actuation cycle. Thirty-two snapshots were systematically acquired during each oscillation period, providing adequate resolution to compute power spectral densities and probability distribution functions across all actuation phases. The phase-wise decomposition mesh utilized for the statistical analysis exhibits substantially refined resolution compared to the snapshot sampling, for simulations conducted with both Mesh $S_1$ and Mesh $S_2$. This furnishes the phase-resolved interrogation essential for describing the physical processes governing the evolution of turbulence statistics, including drag and heat transfer, within each portion of the periodic actuation cycle.

\subsection{Results for optimal drag reduction}

\begin{figure}[h!]
\centering
\includegraphics[scale=0.75]{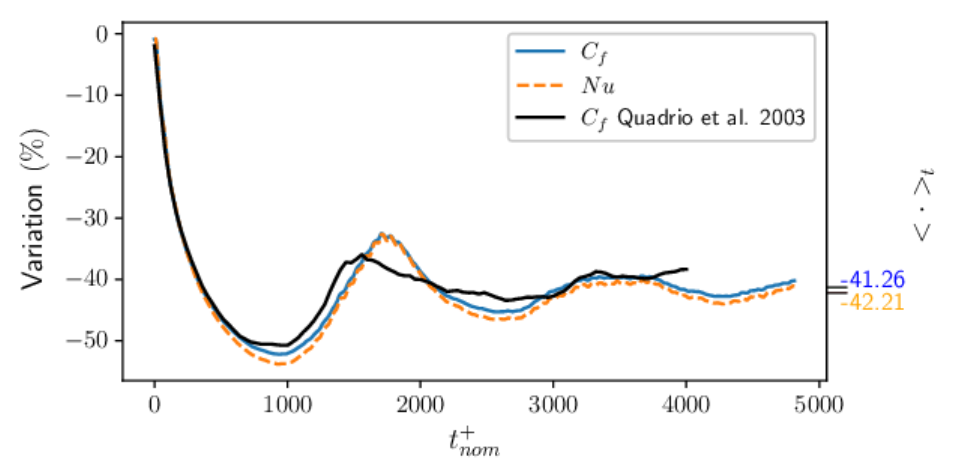}\
\caption{Temporal evolution of drag and heat transfer reductions obtained by spanwise oscillatory wall actuation for control parameters $T^+_{nom}= 125$ and $W^+_{nom}= 18$. The black curve denotes data from \citet{drag3} for reference.}
\label{fig:DR_T125}
\end{figure}

Substantial research within the aerodynamics field has focused on leveraging spanwise wall oscillation for drag reduction and ultimately energy savings. A recent comprehensive review  \citep{2} summarizes these efforts for interested readers. Parametric studies at low Reynolds numbers have identified optimal dimensionless oscillation amplitude ($W^+$) and period ($T^+$) that maximize drag decrease. Significantly, \citet{drag3} demonstrated approximately 40\% drag reduction (after the transient phase) at $\Re_\tau=200$  utilizing nominal control parameters of $T^+_{nom}=125$ and $W^+_{nom} =18$. In figure \ref{fig:DR_T125}, the temporal evolution of the drag reduction obtained by \citet{drag3} is represented by the black line. The blue line depicts the analogous temporal drag variation, defined as:

\begin{equation}
\label{eq:var}
\text{Variation} = \left(\frac{\cfa{ac}}{\langle\cfa{nom} \rangle_t} -1\right) \times 100,
\end{equation}
acquired in the present study, implementing identical oscillation parameters, albeit at a marginally lower Reynolds number. The obtained trend closely resembles that of \citet{drag3}, with the drag reduction converging to $\sim40\%$, validating the implemented wall actuation methodology.

Additional validation simulations were conducted using oscillation parameters of $T^+_{nom}= 200$ and $W^+_{nom}=12$ on the $S2$ mesh. The resulting mean drag reduction exhibited a 2.10\% deviation compared to the reference data of \citet{drag4} (not shown here). This minor deviation is statistically inconsequential and could potentially arise from lack of convergence or the documented Reynolds number effects described by \citet{leschziner2012}. Specifically, they demonstrated that actuation efficacy diminishes with increasing $\Re_\tau$.

In figure \ref{fig:DR_T125}, the temporal evolution of the Nusselt number is also shown, displaying strong agreement with the drag variation. Similar to the drag reduction, approximately 40\% decrease in heat transfer is reached. These results underscore that, comparable to drag, spanwise wall oscillation can substantially reduce heat transfer. The tight correlation between modulated heat transfer and drag is consistent with previous results obtained by  \citet{Fang2010_active}. However, it should be noted that these latters were obtained for a compressible flow using LES methods and a more complex form of wall motion referred to as out-of-phase and in-phase active spanwise wall fluctuations. 

\section{Enhancing heat transfer}
\label{seq:Results}
Since short-period oscillations disrupt streak formation \citep{drag2,leschziner2012}, the reduced drag and heat transfer matches expectations as weakened streaks lead to weaker ejection and sweeping events, reducing mixing momentum. However, as oscillation periods lengthen, there is potential for streak enhancement which could instead intensify turbulent transport \citep{Wenjun_et_al_2019}, increasing the drag and potentienly the heat transfer as well.

Conventional spanwise oscillation control has overwhelmingly focused on drag reduction, with less emphasis on modulating heat transfer. However, the coupled nature of heat and momentum transport implies oscillation parameters increasing drag may also intensify heat transfer. This work deliberately examines one such drag-amplifying actuation scenario to fundamentally characterize the accompanying heat transfer response. As reviewed in section \ref{seq:intro}, the sole documented instance of drag amplification at low Reynolds numbers utilized oscillation period $T^+_{nom}=500$ \citep{drag6}. Given the previously observed tight coupling between drag and heat transfer responses expected from Reynolds analogy, analogous heat transfer enhancement may arise for such oscillatory parameters.

Accordingly, oscillation parameters ${T^+_{nom}=500 \ \mathrm{and} \ W^+_{nom}=30}$ were selected herein to determine if the Reynolds analogy holds for drag increase, wherein heat transfer and drag augmentation occur concurrently. Alternatively, the objective is to ascertain if this analogy can be broken, achieving disproportionately higher heat transfer enhancement exceeding the drag increase. Fundamentally, this work aims to elucidate if heat transfer can be increased and the potential decoupling of heat and momentum transport achieved through spanwise oscillatory actuation.

While disproportionate enhancement of heat transfer over drag is desirable, determining optimal parameters to maximize this dissimilarity exceeds the scope of the present study. Rather, the focus is restricted to fundamental characterization of the heat transfer increase and analogy breaking mechanisms induced by the oscillatory wall forcing.

The primary objective is quantifying the resultant impacts on drag and heat transfer, with particular interest in determining if attainable heat transfer enhancement exceeds the drag increase.

\subsection{Dissimilar heat transfer}

\begin{figure}
    \centering
    \includegraphics[scale=0.7]{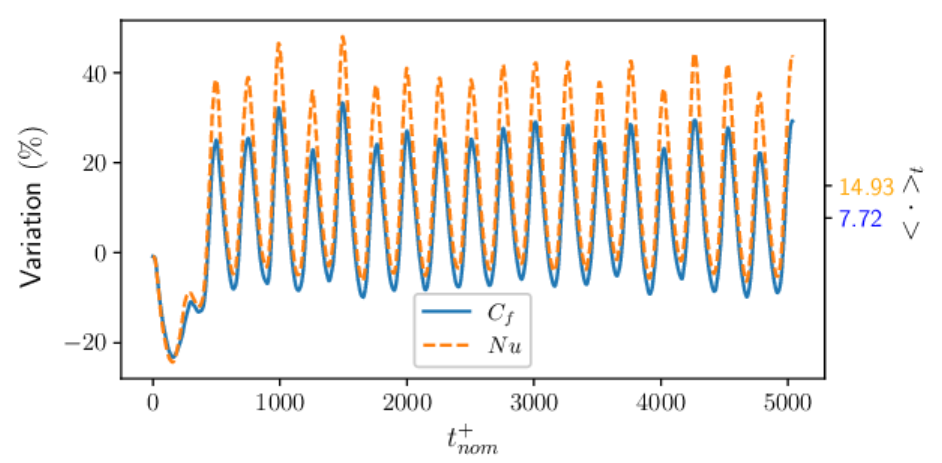}
    \caption{Temporal evolution of drag and heat transfer augmentations obtained by spanwise oscillatory wall actuation for control parameters $T^+_{nom}= 500$ and $W^+_{nom}= 30$.}
    \label{fig:DR_T500_$S_1$}
  \end{figure}

The temporal evolution of the spatially-averaged  drag and heat transfer when applying spanwise wall oscillation with control parameters $T^+_{nom}=500$ and $W^+_{nom}=30$ is conveyed by figure \ref{fig:DR_T500_$S_1$}.

Both heat transfer and skin friction exhibit substantial reduction over the first half oscillation cycle, decreasing by up to $25\%$. This initial attenuation upon actuation commencement is consistent with prior observations, and may be primarily attributed to gradual penetration of the oscillatory Stokes layer into the viscous sublayer and buffer region, disrupting the streak formation process. Thereafter, the heat transfer and friction exhibit periodic oscillations at the actuation frequency of $T^+_{nom}=500$. The friction drag varies between $5\%$ below and $30\%$ above the baseline, averaging $7.72\%$ enhancement. In contrast, the heat-transfer minima are marginally higher while the maxima surge to approximately $45\%$ above baseline. This disproportionate heat-transport amplification yields $15\%$ averaged thermal intensification, doubling the friction increase. Overall, the results underscore the efficacy of spanwise oscillations in elevating turbulent heat convection. More significantly, they expose pronounced asymmetry between modulated heat transfer and drag, the heat transfer oscillations possess substantially greater amplitudes and gains compared to friction fluctuations.

  \begin{figure}
  \centering
    \centering
    \includegraphics[scale=0.7]{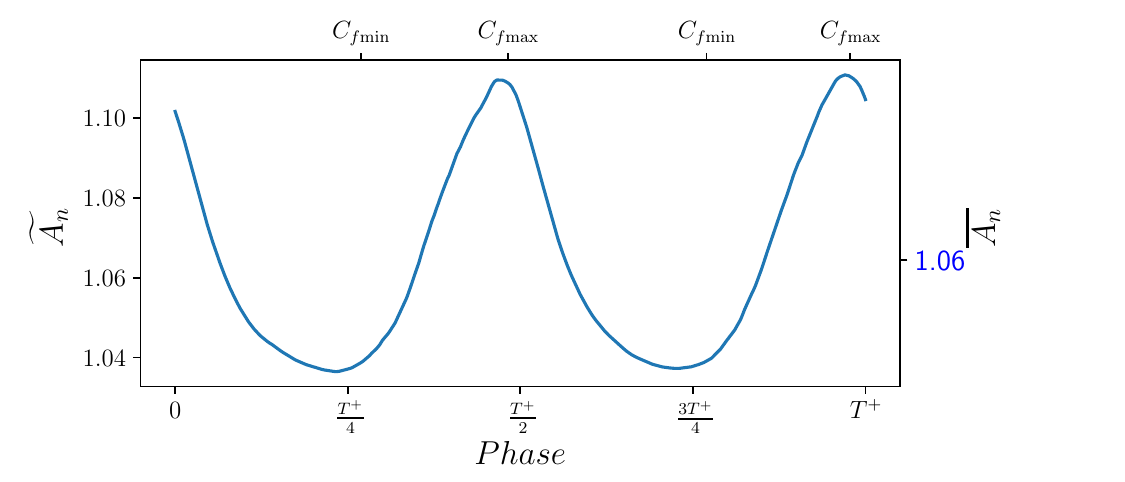}
    \caption{Phase-wise averaged evolution of the analogy factor $\widetilde{A_n}$ obtained by spanwise oscillatory wall actuation for control parameters $T^+_{nom}= 500$ and $W^+_{nom}= 30$.}
    \label{fig:An_T500_$S_1$}
  \end{figure}

An analogy factor $A_n$ is introduced to quantify differences between heat transfer and drag responses, as previously performed by \citet{DHT2}. This factor, defined as the ratio of Nusselt number to friction coefficient, is given by

\begin{equation}
    \label{eq:An}
    A_n = \frac{\Nu_{ac}/\Nu_{nom}}{\cfa{ac}/\cfa{nom}}.
\end{equation}
Control parameters resulting in $A_n > 1$ indicate greater heat-transfer enhancement relative to the accompanying drag increase.

Through the phase-averaged analogy factor $\widetilde{A_n}$, figure \ref{fig:An_T500_$S_1$} distinctly shows the pronounced friction and heat-transfer asymmetry over oscillation phases. $\widetilde{A_n}$ reaches a minimum of slightly below $1.04$ during friction minima. Meanwhile, it strengthens to $1.11$ coinciding with peak friction coefficient. The shorter duration of $\widetilde{A_n}$ intensification compared to attenuation might indicate that distinct mechanisms drive the reduction and augmentation stages, and also lead to an average value of $\widetilde{A_n}$ equal to $1.06$, highlighting a greater heat-transfer enhancement.

At $\Pr=1$, a comparison can be made to the streamwise travelling wave-like wall deformation used in \citet{DHT2} at $\Re_\tau=180$ with constant temperature difference (CTD) boundary conditions. They obtained an average analogy factor of $\overline{A_n} =1.13$ with optimal parameters. Even though, the control method is different, a comparison with the current study can be performed considering the similarities in Reynolds and Prandtl numbers. The current results using spanwise oscillations give a lower time-averaged value of $\overline{A_n} =1.06$ with mixed boundary conditions. This reduced performance is likely because dissimilarity is more difficult to achieve using mixed versus CTD conditions, as shown in \citet{DHT4}. Additionally, only one parameter set was tested here, so higher analogy factors may be achievable by optimizing the parameters. Nevertheless, the mixed boundary conditions pose an inherent challenge for maximizing dissimilarity that steady CTD does not encounter.
 
\subsection{FIK Identity Component Analysis of Transport Phenomena}
This section aims to shed the light on the mechanisms resulting in the breakdown of the Reynolds analogy under imposed spanwise wall oscillation. To this end, the impact of the forcing terms appearing in the continuous equations is examined. A secondary objective is to attain a deeper understanding of the link between frictional quantities and turbulent stresses. In pursuit of these aims, an intricate dissection of the friction coefficient and Nusselt/Stanton numbers into discrete components will be orchestrated through the utilization of the FIK identities, as delineated in \citet{drag5}, and their thermal analog presented in \citet{DHT5}. 

 \begin{equation}
\label{eq:FIK_cf_nu}
\begin{aligned}
&\overline{\Nu} = 
\underbrace{6}_{\textcolor{blue}{\Nu_l}} +  \underbrace{ 3\Pe\int_{0}^{2} (y-1) \overline{\Theta^{\prime\prime}v^{\prime\prime}} \diff y}_{\textcolor{mygreen}{\Nu_t}}     
-  \underbrace{\frac{3\Pe A}{2}   \int_{0}^2  (y-1)^2(\overline{u}_b -  \overline{u}) \diff y}_{\textcolor{red}{\Nu_f}} \\
&\widetilde{\Nu} = \overline{\Nu} + \underbrace{ 3\Pe \int_{0}^{2} (y-1) \reallywidehat{\Theta^{\prime \prime}v^{\prime \prime}} \diff y}_{\textcolor{orange}{\Nu_p}} \\
&\overline{\cfa{}} = \underbrace{\frac{9}{\Re}}_{\textcolor{blue}{\cfa{l}}} + \underbrace{\frac{27}{4} \int_{0}^{2} (y-1) \overline{u^{\prime\prime}v^{\prime\prime}} \diff y}_{\textcolor{mygreen}{\cfa{t}}}   \\
& \widetilde{\cfa{}}=\overline{\cfa{}} + \underbrace{\frac{27}{4} \int_{0}^{2} (y-1) \reallywidehat{u^{\prime \prime}v^{\prime \prime}} \diff y}_{\textcolor{orange}{\cfa{p}}} 
\end{aligned} \tag{5.2}
\end{equation}
 
By analyzing each term of the decomposition individually, a more sophisticated comprehension of the physical quantities directly related to drag/heat transfer variations is anticipated. It is known that the friction coefficient $\widetilde{\cfa{}}$ is directly linked to the turbulent shear stress $\widetilde{u''v''}$, and that the Nusselt number is connected to the turbulent heat flux $\widetilde{\Theta''v''}$. This link will be clarified through the following analysis. It is expected that the more pronounced enhancement in heat transfer should also be discernible when comparing the turbulent shear stress and heat flux.

 Adapting the FIK identities to the current configuration reveals the decompositions obtained for the Nusselt number and the friction coefficient averaged in time and in phase, which are detailed in equation \eqref{eq:FIK_cf_nu}.

The initial component ($\cfa{l}$ or $\Nu_l$) is referred to as the laminar component, while the second ($\cfa{t}$ or $\Nu_t$) represents the time wise turbulent component and the third ($\cfa{p}$ or $\Nu_p$) is referred to as the turbulent phase wise component. As far as the Nusselt number is concerned, the fourth term ($\Nu_f$) is the source term component, which is absent from the decomposition of the friction coefficient because of its negligible value. These formulations establish and clarify the direct relationship between drag/heat transfer and turbulent shear stress/heat flux. In addition to distinguishing between turbulent components averaged in time and phase, these decompositions also elucidate the contribution of forcing terms to drag/heat transfer.

\begin{figure}[h!]
  \centering
     \includegraphics[trim= 100 50 100  120, clip=true,scale=0.21]{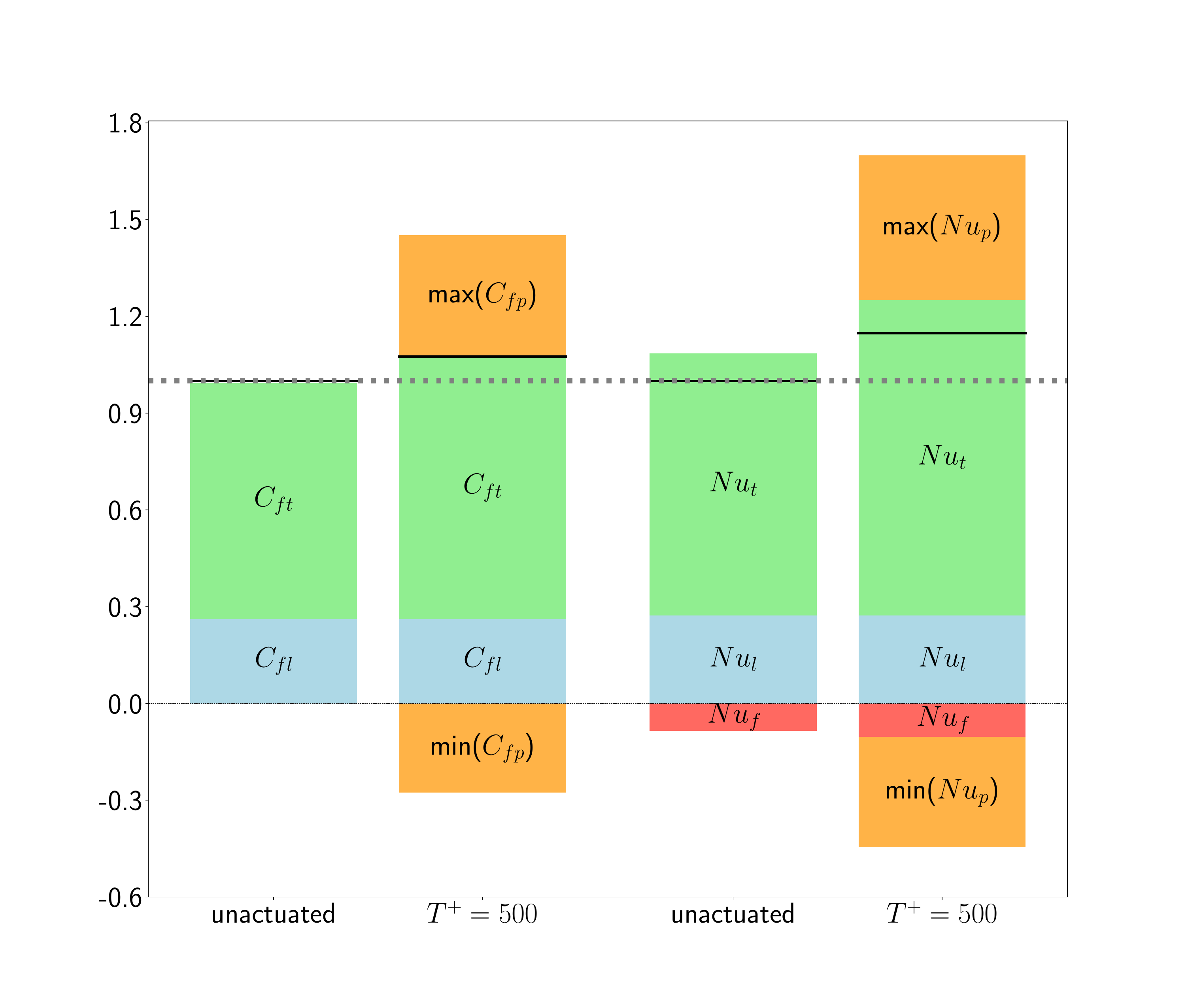}  
  \caption{FIK identity contributions of $\cf$ and $Nu$ (normalized by $\cf_{nom}$ and $Nu_{nom}$). Blue bins correspond to the laminar component, green bins to the turbulent time-wise component, red bins to the forcing component and orange bins to the minimum and maximum of the turbulent phase-wise component. \ \protect\greydots \ : \ total unactuated value, \ \protect\blackline \ : \ total value.}
  \label{fig:FIK}
\end{figure}

Figure \ref{fig:FIK} shows the results derived from the FIK identity decomposition of the friction coefficient and Nusselt number under conditions of $T^+_{nom}=500$ and $W^+_{nom}=30$, relative to the unactuated scenario. Each component represented is normalised with respect to the total unactuated value. The greater phase variations observed in the upper orange bins, compared with the lower bins, highlights a more pronounced difference between the fluctuations of the average and high intensity stresses, compared with the difference in fluctuations between the low intensity and average values. Negative values of the $\Nu_f$ component weaken heat transfer, as shown by the red bins. However, the small differences between actuated and unactuated $\Nu_f$ suggest that the forcing term has a negligible effect on increasing heat transfer. It should be noted that, as stated in section \ref{seq:config}, the only differences in the continuous equations governing streamwise velocity and temperature are the source terms, the solenoidality of the velocity and the linearity of the convective term in the temperature equation.

In particular, the negligible impact of the forcing term has been established with respect to dissimilarity, meaning the disparities observed between $\langle \frac{\Nu_{ac}}{\Nu_{nom}} \rangle_{t}$ and $\langle \frac{\cfa{ac}}{\cfa{nom}} \rangle_{t}$  are attributed to the time-dependent turbulent components of the FIK identities. This dissimilarity is therefore due to discrepancies between the actuated turbulent shear stress and turbulent heat flux. Furthermore, given the demonstrated insignificance of the forcing term, the differences in the actuated drag and heat transfer increases arise from the solenoidal velocity condition and/or the linear temperature equation.

\begin{figure}[h!]
\centering
\subfigure[]{\label{fig:uvprime_nom}\includegraphics[trim= 0 0 0  0, clip=true,width=0.44\textwidth]{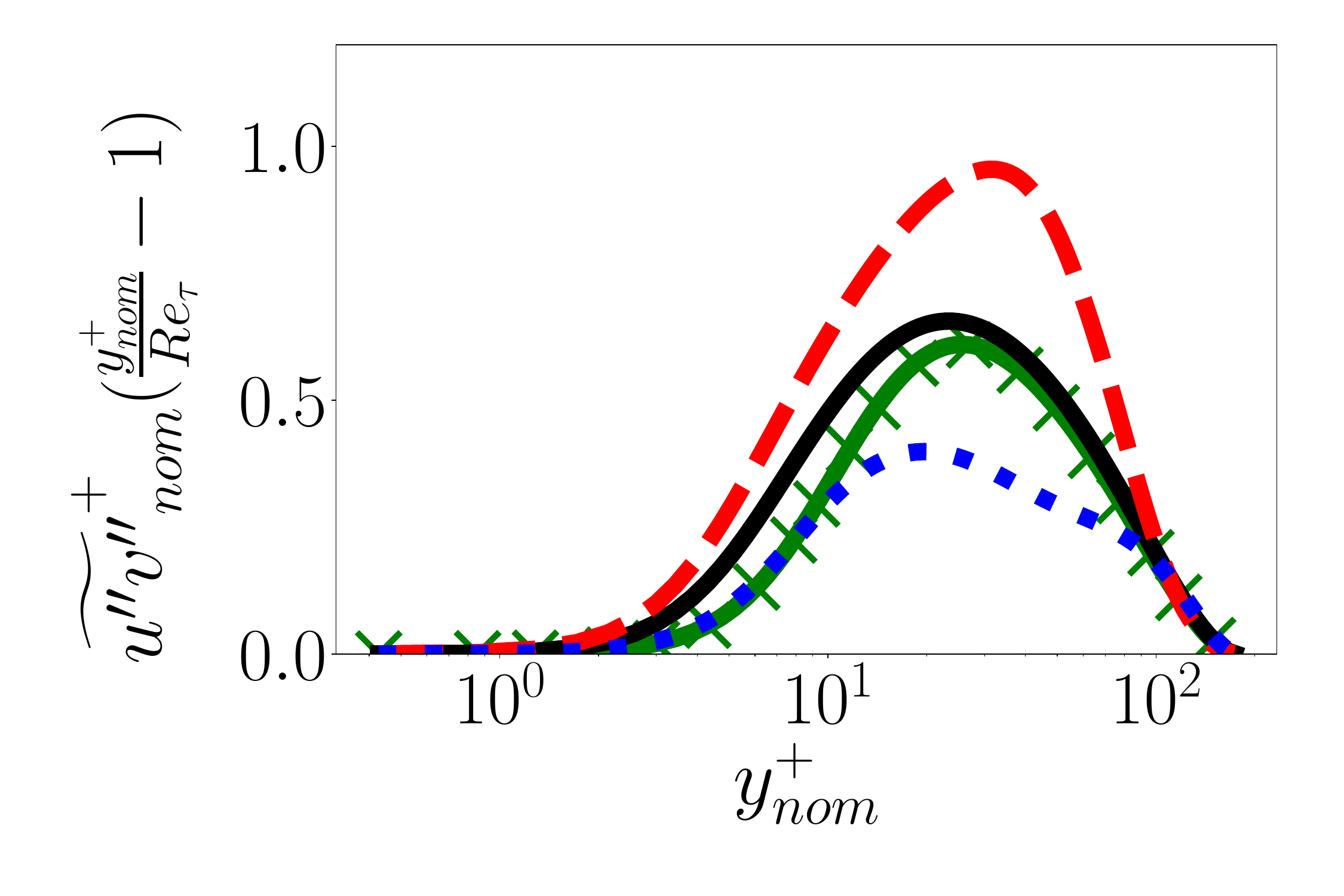}}
\subfigure[]{\label{fig:vthetaprime_nom}\includegraphics[trim= 0 0 0  0, clip=true,width=0.44\textwidth]{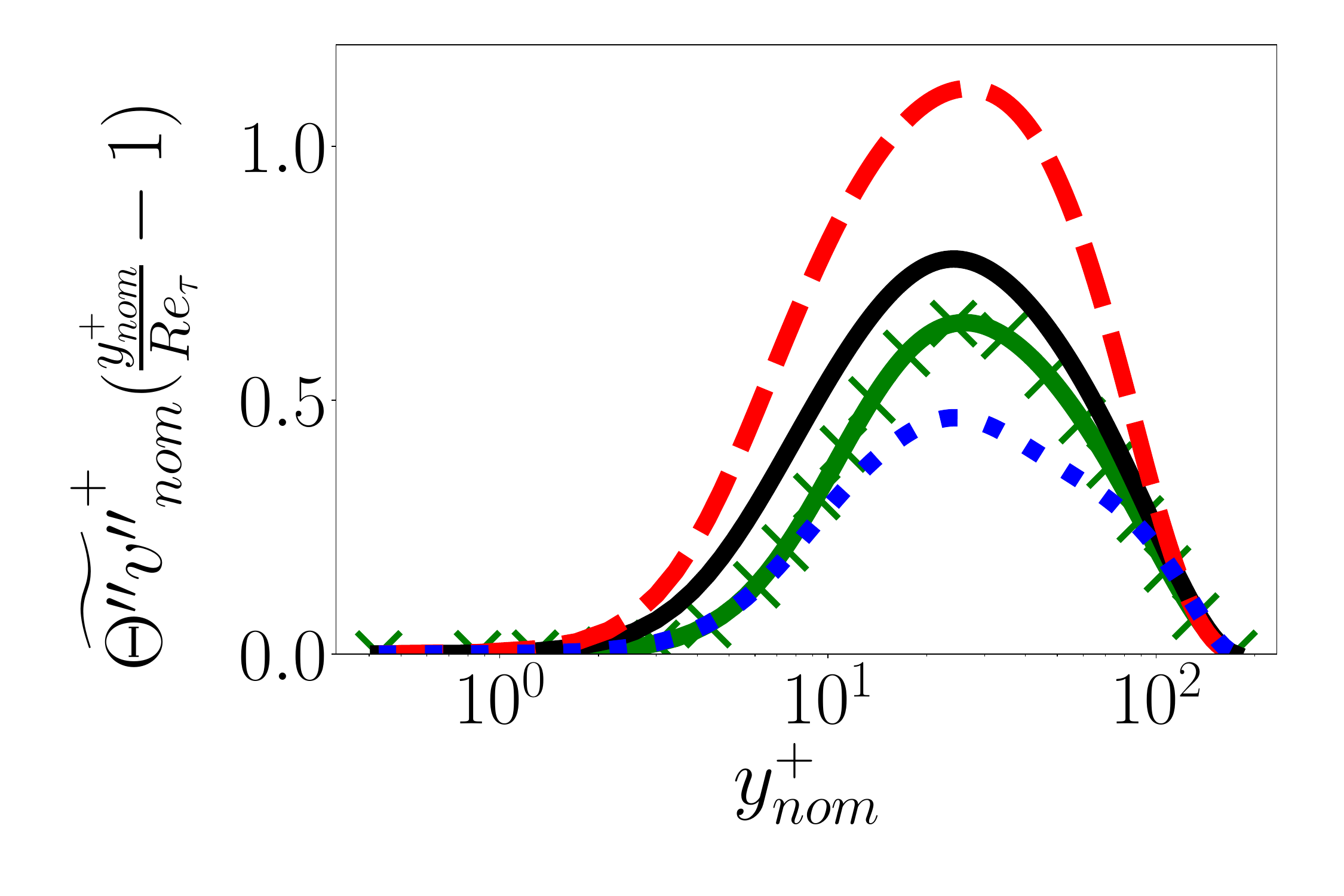}}
\caption{Wall normal distributions on mesh $S_2$ of (a): $\widetilde{u^{\prime \prime}  v^{\prime \prime}}^+_{nom}(\frac{y^+_{nom}}{\Re_\tau}-1)$, (b): $\widetilde{\Theta^{\prime \prime} v^{\prime \prime}}^+_{nom}(\frac{y^+_{nom}}{\Re_\tau}-1)$. \ \protect\bluedots \ : \  actuated stress on $\cfa{\mathrm{min}}$ phases, \protect\reddashed \ : \ on  $\cfa{\mathrm{max}}$ phases, \\ \protect\blackline \ : \ actuated time-wise stress, \ \protect\greenbar \ : \ unactuated stress.}
 \label{fig:stat_ac}
\end{figure}

The turbulent components of the FIK identities (equation \eqref{eq:FIK_cf_nu}) contain a weighting function of $y- 1$, which weights the shear stresses as a function of proximity to the wall. Shear stresses close to the wall thus bring a greater contribution than those closer to the centre of the channel. Therefore, strategies that alter the magnitude or very near-wall distribution of turbulent shear stress and heat flux can have a significant impact on drag and heat transfer. For example, differential thinning of the thermal sublayer relative to the viscous sublayer could explain the observed dissimilarity in drag and heat transfer enhancement.

Figures \ref{fig:uvprime_nom} and \ref{fig:vthetaprime_nom} show the wall-normal distribution of the weighted actuated turbulent shear stress and heat flux, respectively. The red line corresponds to the maximum drag phases, the blue line to the minimum drag phases and the black line represents the time average, while the green lines, for reference, show the time-averaged distributions obtained for the unactuated case.

For the shear stress, it is clear that the increase in drag is due to the actuation-induced increase in the weighted turbulent shear stress near the wall, which extends to the peak. This influence on drag is particularly pronounced, given the weight function in the turbulent terms of the FIK identities. A similar conclusion can be drawn for the turbulent heat flux, albeit with a more substantial increase in the weighted stress at the peak. As expected, the dissimilarity in between the increases in drag and heat transfer becomes  more apparent when examining the profiles of turbulent shear stress and heat flux, which results in the enhancement in heat transfer exceeding that of drag, leading to the observed dissimilarity with $\overline{A_n}=1.06$.
A detailed comparison between the $\cfa{\mathrm{min}}$ and $\cfa{\mathrm{max}}$ phases highlights once more the importance of  the weighted stresses' influence on the simultaneous increase in drag and heat transfer.
In addition, this analysis reaffirms the observation that phases marked by pronounced fluctuations in velocity and temperature are further away from time-averaged quantities.

\section{Conclusion}
This study investigated the use of spanwise wall oscillation to control both drag and heat transfer. Direct numerical simulations at $Re_{\tau}\approx180$ and $\Pr=1$ demonstrated that oscillation parameters known to reduce drag ($T^+=125$, $W^+=18$) also decreased heat transfer in a similarly coupled manner, maintaining the close connection between momentum and thermal transport. The preservation of the Reynolds analogy was anticipated, as past work has shown that for these control parameters, the Stokes layer induced by the wall oscillations dampens the streaks and associated ejection-sweep motions that drive mixing momentum. Therefore, it is unsurprising that the weakened mixing simultaneously reduces both drag and convective heat transport by similar margins.

Conversely, intensified momentum mixing is anticipated to amplify convective heat transfer. This expectation was validated in the current study through a direct numerical simulation with a large period of $T^+=500$ and amplitude of $W^+=30$. More remarkably, the heat transfer strengthened substantially more than the drag, as the Nusselt number rose approximately a couple times more than the friction coefficient, increasing by 15\% and 7.7\%, respectively. The more substantial amplification of heat transfer compared to drag underscores the prospect of selectively magnifying thermal convection, breaking the Reynolds analogy, using spanwise wall motion. The current study shows how dissimilarity can be achieved by enhancing heat transfer more than drag, by controlling near-wall turbulence, however, conducting parametric optimization to maximise this dissimilarity was out of the scope of the present study. 

A first insight into the physical mechanisms behind the actuation effect is obtained by using the FIK identity decomposition for the coefficient of friction and the Nusselt number. The FIK identity analysis shed light on the physical mechanisms governing the observed heat transfer enhancement exceeding drag rise under spanwise wall oscillation. The negligible impact of the forcing terms showed that the perceived dissimilarity arises from the divergence-free velocity condition and linear temperature equation.  To strengthen the dissimilarity, the turbulent heat flux has to be amplified over the turbulent shear stress in the near wall region. The present study demonstrates this can be achieved for spanwise wall oscillation with period $T^+=500$ and amplitude $W^+=30$, where a more substantial relative increase occurs in the heat flux $\overline{\Theta^{\prime}v^{\prime}}$ compared to the shear stress $\overline{u^{\prime}v^{\prime}}$. This investigation explains how preferential strengthening of the near-wall heat flux can disrupt the Reynolds analogy, selectively magnifying heat convection through intentional manipulation of wall-turbulent structure.  Further probing of the fundamental dynamics and optimization of parameters may uncover approaches for maximizing the dissimilarity between momentum and thermal transport, improving efficiency.

\section*{Declaration of interest}
The authors report no conflict of interest.

\section*{Acknowledgements}
We express our sincere gratitude to Michael Leschziner, emeritus professor at Imperial College London, for his invaluable insights and guidance, which proved to be instrumental throughout the course of this research. 
This work was performed using HPC resources from GENCI-IDRIS (Grant 2023-gen14284)

This work pertains to the French government program "Investissements d'Avenir" (EUR INTREE, reference ANR-18-EURE-0010, and LABEX INTERACTIFS, reference ANR-11-LABX-0017-01).

This work was supported by the CPER-FEDER project of Region Nouvelle Aquitaine.

\section*{Fundings}

Our research activities are supported by the French Agence Nationale de la Recherche (ANR) in
the framework of the project ”Apprentissage automatique pour les
récepteurs solaires à haute température – SOLAIRE” (ANR-21-CE50-0031).

\section*{Data availability statement}

The data that support the findings of this study are available from the corresponding author, Lou Guérin, upon reasonable request.


\begin{thebibliography}{45}
\expandafter\ifx\csname natexlab\endcsname\relax\def\natexlab#1{#1}\fi
\def\au#1{#1} \def\ed#1{#1} \def\yr#1{#1}\def\at#1{#1}\def\jt#1{\textit{#1}} \def\bt#1{#1}\def\bvol#1{\textbf{#1}} \def\vol#1{#1} \def\pg#1{#1} \def\publ#1{#1}\def\arxiv#1{#1}\def\org#1{#1}\def\st#1{\textit{#1}}

\bibitem[Abdulbari {\em et~al.\/}(2013)Abdulbari, Zulkifli, Salleh \& Yusoff]{7}
{\sc \au{Abdulbari, H.A.}, \au{Zulkifli, R.}, \au{Salleh, M.Z.} \& \au{Yusoff, M.Z.}} \yr{2013}  \at{Skin-friction drag reduction using riblets, dimples, and oscillating wall}.  \jt{Journal of Mechanical Engineering and Sciences}  \bvol{4},  \pg{452--464}.

\bibitem[Abe {\em et~al.\/}(2009)Abe, Antonia \& Kawamura]{thermo8}
{\sc \au{Abe, H.}, \au{Antonia, R.~A.} \& \au{Kawamura, H.}} \yr{2009}  \at{Correlation between small-scale velocity and scalar fluctuations in a turbulent channel flow}.  \jt{Journal of Fluid Mechanics}  \bvol{627},  \pg{1–32}.

\bibitem[Agostini \& Leschziner(2021)]{agostini2021statistical}
{\sc \au{Agostini, Lionel} \& \au{Leschziner, Michael}} \yr{2021}  \at{Statistical analysis of outer large-scale/inner-layer interactions in channel flow subjected to oscillatory drag-reducing wall motion using a multiple-variable joint-probability-density function methodology}.  \jt{Journal of Fluid Mechanics}  \bvol{923},  \pg{A25}.

\bibitem[Agostini {\em et~al.\/}(2017)Agostini, Leschziner, Poggie, Bisek \& Gaitonde]{agostini2017multi}
{\sc \au{Agostini, Lionel}, \au{Leschziner, Michael}, \au{Poggie, Jonathan}, \au{Bisek, Nicholas~J} \& \au{Gaitonde, Datta}} \yr{2017}  \at{Multi-scale interactions in a compressible boundary layer}.  \jt{Journal of Turbulence}  \bvol{18}~(8),  \pg{760--780}.

\bibitem[Agostini {\em et~al.\/}(2014{\natexlab{{\em a\/}}})Agostini, Touber \& Leschziner]{22}
{\sc \au{Agostini, L.}, \au{Touber, E.} \& \au{Leschziner, M.A.}} \yr{2014{\natexlab{{\em a\/}}}}  \at{Spanwise oscillatory wall motion in channel flow: Drag-reduction mechanisms inferred from dns-predicted phase-wise property variations}.  \jt{Journal of Fluid Mechanics}  \bvol{743},  \pg{606--635}.

\bibitem[Agostini {\em et~al.\/}(2014{\natexlab{{\em b\/}}})Agostini, Touber \& Leschziner]{drag1}
{\sc \au{Agostini, L.}, \au{Touber, E.} \& \au{Leschziner, M.}} \yr{2014{\natexlab{{\em b\/}}}}  \at{The turbulence vorticity as a window to the physics of friction-drag reduction by oscillatory wall motion}.  \jt{International Journal of Heat and Fluid Flow}  \bvol{51}.

\bibitem[Agostini {\em et~al.\/}(2014{\natexlab{{\em c\/}}})Agostini, Touber \& Leschziner]{drag2}
{\sc \au{Agostini, L.}, \au{Touber, E.} \& \au{Leschziner, M.~A.}} \yr{2014{\natexlab{{\em c\/}}}}  \at{Spanwise oscillatory wall motion in channel flow: drag-reduction mechanisms inferred from dns-predicted phase-wise property variations at re tau =1000}.  \jt{Journal of Fluid Mechanics}  \bvol{743},  \pg{606–635}.

\bibitem[Alcántara(2022)]{thermo1}
{\sc \au{Alcántara, F.}} \yr{2022}  \at{Study of the thermal field of turbulent channel flows via direct numerical simulations}. PhD thesis.

\bibitem[Asidin {\em et~al.\/}(2019)Asidin, Kuntjoro \& Salleh]{5}
{\sc \au{Asidin, S.}, \au{Kuntjoro, W.} \& \au{Salleh, M.Z.}} \yr{2019}  \at{A review on flow control methods in turbine internal cooling}.  \jt{Applied Sciences}  \bvol{9}~(16),  \pg{3233}.

\bibitem[Bartholomew {\em et~al.\/}(2020)Bartholomew, Deskos, Frantz, Schuch, Lamballais \& Laizet]{Xcompact3}
{\sc \au{Bartholomew, P.}, \au{Deskos, G.}, \au{Frantz, R.~A.S.}, \au{Schuch, F.~N.}, \au{Lamballais, E.} \& \au{Laizet, S.}} \yr{2020}  \at{Xcompact3d: An open-source framework for solving turbulence problems on a cartesian mesh}.  \jt{SoftwareX}  \bvol{12},  \pg{100550}.

\bibitem[Campo(2019)]{thermo7}
{\sc \au{Campo, G. F.~Narváez}} \yr{2019}  \at{Fluid-solid thermal coupling in pipe and channel turbulent flows via a dual immersed boundary method}. PhD thesis, Universidade Federal do Rio Grande do Sul.

\bibitem[Choi {\em et~al.\/}(1998)Choi, DeBisschop \& Clayton]{21}
{\sc \au{Choi, K.S.}, \au{DeBisschop, J.R.} \& \au{Clayton, B.R.}} \yr{1998}  \at{Turbulent boundary-layer control by means of spanwise-wall oscillation}.  \jt{AIAA Journal}  \bvol{36}~(7),  \pg{1157--1162}.

\bibitem[Choi \& Clayton(2001)]{18}
{\sc \au{Choi, Kwing-So} \& \au{Clayton, B.R.}} \yr{2001}  \at{The mechanism of turbulent drag reduction with wall oscillation}.  \jt{International Journal of Heat and Fluid Flow}  \bvol{22},  \pg{1--9}.

\bibitem[Cruz(2021)]{thermo5}
{\sc \au{Cruz, R.~Vicente}} \yr{2021}  \at{High-fidelity simulation of conjugate heat transfer between a turbulent flow and a duct geometry}. PhD thesis.

\bibitem[Fang {\em et~al.\/}(2011)Fang, Li, Cambonie, Lebey \& Lu]{Fang2011}
{\sc \au{Fang, H.}, \au{Li, Q.}, \au{Cambonie, T.}, \au{Lebey, M.} \& \au{Lu, L.}} \yr{2011}  \at{Two-point correlation analysis of heat transfer in an oscillating channel flow}.  \jt{International Journal of Heat and Mass Transfer}  \bvol{54}~(9-10),  \pg{2016--2026}.

\bibitem[Fang \& Lu(2010)]{Fang2010_active}
{\sc \au{Fang, J.} \& \au{Lu, L.}} \yr{2010}  \at{Large eddy simulation of compressible turbulent channel flow with active spanwise wall fluctuations}.  \jt{Modern Physics Letters B - MOD PHYS LETT B}  \bvol{24},  \pg{1457--1460}.

\bibitem[Fang {\em et~al.\/}(2009)Fang, Lu \& Shao]{Fang2009}
{\sc \au{Fang, J.}, \au{Lu, L.} \& \au{Shao, L.}} \yr{2009}  \at{Large eddy simulation of compressible turbulent channel flow with spanwise wall oscillation}.  \jt{Science in China Series G: Physics, Mechanics and Astronomy}  \bvol{52},  \pg{1233--1243}.

\bibitem[Fang {\em et~al.\/}(2010)Fang, Lu \& Shao]{Fang2010}
{\sc \au{Fang, J.}, \au{Lu, L.} \& \au{Shao, L.}} \yr{2010}  \at{Heat transport mechanisms of low mach number turbulent channel flow with spanwise wall oscillation}.  \jt{Acta Mechanica Sinica}  \bvol{26},  \pg{391--399}.

\bibitem[Flageul(2015)]{thermo10}
{\sc \au{Flageul, C.}} \yr{2015}  \at{Refined database by dns for turbulence effects on near-wall heat transfer}. PhD thesis.

\bibitem[Flageul {\em et~al.\/}(2015)Flageul, Benhamadouche, Lamballais \& Laurence]{thermo6}
{\sc \au{Flageul, C.}, \au{Benhamadouche, S.}, \au{Lamballais, E.} \& \au{Laurence, D.}} \yr{2015}  \at{Dns of turbulent channel flow with conjugate heat transfer: Effect of thermal boundary conditions on the second moments and budgets}.  \jt{International Journal of Heat and Fluid Flow}  \bvol{55},  \pg{34--44}, special Issue devoted to the 10th Int. Symposium on Engineering Turbulence Modelling and Measurements (ETMM10) held in Marbella, Spain on September 17-19, 2014.

\bibitem[Gatti \& Quadrio(2013)]{Gatti2013}
{\sc \au{Gatti, D.} \& \au{Quadrio, M.}} \yr{2013}  \at{Performance losses of drag-reducing spanwise forcing at moderate values of the reynolds number}.  \jt{Physics of Fluids}  \bvol{25}~(12).

\bibitem[Gatti \& Quadrio(2016)]{Gatti2016}
{\sc \au{Gatti, D.} \& \au{Quadrio, M.}} \yr{2016}  \at{Reynolds-number dependence of turbulent skin-friction drag reduction induced by spanwise forcing}.  \jt{Journal of Fluid Mechanics}  \bvol{802},  \pg{553--582}.

\bibitem[Gomez {\em et~al.\/}(2009)Gomez, Flutet \& Sagaut]{drag5}
{\sc \au{Gomez, T.}, \au{Flutet, V.} \& \au{Sagaut, P.}} \yr{2009}  \at{Contribution of reynolds stress distribution to the skin friction in compressible turbulent channel flows}.  \jt{Phys. Rev. E}  \bvol{79},  \pg{035301}.

\bibitem[Hasegawa \& Kasagi(2011)]{DHT5}
{\sc \au{Hasegawa, Y.} \& \au{Kasagi, N.}} \yr{2011}  \at{Dissimilar control of momentum and heat transfer in a fully developed turbulent channel flow}.  \jt{Journal of Fluid Mechanics}  \bvol{683},  \pg{57–93}.

\bibitem[Hurst {\em et~al.\/}(2014)Hurst, Yang \& Chung]{Hurst2014}
{\sc \au{Hurst, E.}, \au{Yang, Q.} \& \au{Chung, Y.M.}} \yr{2014}  \at{The effect of reynolds number on turbulent drag reduction by streamwise travelling waves}.  \jt{Journal of Fluid Mechanics}  \bvol{759},  \pg{28--55}.

\bibitem[Jung {\em et~al.\/}(1992)Jung, Mangiavacchi \& Akhavan]{drag6}
{\sc \au{Jung, W.}, \au{Mangiavacchi, N.} \& \au{Akhavan, R.}} \yr{1992}  \at{Suppression of turbulence in wall‐bounded flows by high‐frequency spanwise oscillations}.  \jt{Physics of Fluids A Fluid Dynamics}  \bvol{4}.

\bibitem[Kaithakkal {\em et~al.\/}(2021)Kaithakkal, Kametani \& Hasegawa]{DHT3}
{\sc \au{Kaithakkal, A.}, \au{Kametani, Y.} \& \au{Hasegawa, Y.}} \yr{2021}  \at{Dissimilar heat transfer enhancement in a fully developed laminar channel flow subjected to a traveling wave-like wall blowing and suction}.  \jt{International Journal of Heat and Mass Transfer}  \bvol{164},  \pg{120485}.

\bibitem[Kasagi {\em et~al.\/}(2012)Kasagi, Hasegawa, Fukagata \& Iwamoto]{DHT4}
{\sc \au{Kasagi, N.}, \au{Hasegawa, Y.}, \au{Fukagata, K.} \& \au{Iwamoto, K.}} \yr{2012}  \at{Control of turbulent transport: Less friction and more heat transfer}.  \jt{Journal of Heat Transfer}  \bvol{134}~(3), 031009,  \arxiv{arXiv: https://asmedigitalcollection.asme.org/heattransfer/article-pdf/134/3/031009/5794148/031009\_1.pdf}.

\bibitem[Laizet \& Lamballais(2009)]{Xcompact1}
{\sc \au{Laizet, S.} \& \au{Lamballais, E.}} \yr{2009}  \at{High-order compact schemes for incompressible flows: A simple and efficient method with quasi-spectral accuracy}.  \jt{Journal of Computational Physics}  \bvol{228}~(16),  \pg{5989--6015}.

\bibitem[Laizet \& Li(2011)]{Xcompact2}
{\sc \au{Laizet, S.} \& \au{Li, N.}} \yr{2011}  \at{Incompact3d: A powerful tool to tackle turbulence problems with up to o(105) computational cores}.  \jt{International Journal for Numerical Methods in Fluids}  \bvol{67},  \pg{1735 -- 1757}.

\bibitem[Lamballais {\em et~al.\/}(2011)Lamballais, Fortuné \& Laizet]{LAMBALLAIS20113270}
{\sc \au{Lamballais, E.}, \au{Fortuné, V.} \& \au{Laizet, S.}} \yr{2011}  \at{Straightforward high-order numerical dissipation via the viscous term for direct and large eddy simulation}.  \jt{Journal of Computational Physics}  \bvol{230}~(9),  \pg{3270--3275}.

\bibitem[Marusic {\em et~al.\/}(2021)Marusic, Chandran, Rouhi, Fu, Wine, Holloway, Chung \& Smits]{Marusic2021}
{\sc \au{Marusic, I.}, \au{Chandran, D.}, \au{Rouhi, A.}, \au{Fu, M.K.}, \au{Wine, D.}, \au{Holloway, B.}, \au{Chung, D.} \& \au{Smits, A.J.}} \yr{2021}  \at{An energy-efficient pathway to turbulent drag reduction}.  \jt{Nature Communications}  \bvol{12}~(1),  \pg{5805}.

\bibitem[Ni {\em et~al.\/}(2016)Ni, Lu, Ribault \& Fang]{Ni2016}
{\sc \au{Ni, W.}, \au{Lu, L.}, \au{Ribault, C.~Le} \& \au{Fang, J.}} \yr{2016}  \at{Direct numerical simulation of supersonic turbulent boundary layer with spanwise wall oscillation}.  \jt{Energies}  \bvol{9}~(3).

\bibitem[Quadrio(2011)]{20}
{\sc \au{Quadrio, M.}} \yr{2011}  \at{Drag reduction in turbulent boundary layers by in-plane wall motion}.  \jt{Philosophical Transactions of the Royal Society A: Mathematical, Physical and Engineering Sciences}  \bvol{369}~(1940),  \pg{1428--1442}.

\bibitem[Quadrio \& Ricco(2003)]{drag3}
{\sc \au{Quadrio, M.} \& \au{Ricco, P.}} \yr{2003}  \at{Initial response of a turbulent channel flow to spanwise oscillation of the walls}.  \jt{Journal of Turbulence}  \bvol{4},  \pg{1--23}.

\bibitem[Quadrio \& Ricco(2004)]{drag4}
{\sc \au{Quadrio, M.} \& \au{Ricco, P.}} \yr{2004}  \at{Critical assessment of turbulent drag reduction through spanwise wall oscillations}.  \jt{Journal of Fluid Mechanics}  \bvol{521},  \pg{251–271}.

\bibitem[Ricco {\em et~al.\/}(2021)Ricco, Skote \& Leschziner]{2}
{\sc \au{Ricco, P.}, \au{Skote, M.} \& \au{Leschziner, M.~A.}} \yr{2021}  \at{A review of turbulent skin-friction drag reduction by near-wall transverse forcing}.  \jt{Progress in Aerospace Sciences}  \bvol{123},  \pg{100713}.

\bibitem[Seki {\em et~al.\/}(2006)Seki, Iwamoto \& Kawamura]{thermo9}
{\sc \au{Seki, Y.}, \au{Iwamoto, K.} \& \au{Kawamura, H.}} \yr{2006}  \at{Prandtl number effect on turbulence statistics through high spatial resolution dns of turbulent heat transfer in a channel flow}.  \jt{Transactions of the Japan Society of Mechanical Engineers. B}  \bvol{72},  \pg{2856--2861}.

\bibitem[Touber \& Leschziner(2012)]{leschziner2012}
{\sc \au{Touber, E.} \& \au{Leschziner, M.~A.}} \yr{2012}  \at{Near-wall streak modification by spanwise oscillatory wall motion and drag-reduction mechanisms}.  \jt{Journal of Fluid Mechanics}  \bvol{693},  \pg{150–200}.

\bibitem[Uchino {\em et~al.\/}(2017)Uchino, Mamori \& Fukagata]{DHT2}
{\sc \au{Uchino, K.}, \au{Mamori, H.} \& \au{Fukagata, K.}} \yr{2017}  \at{Heat transfer in fully developed turbulent channel flow with streamwise traveling wave-like wall deformation}.  \jt{Journal of Thermal Science and Technology}  \bvol{12}~(1),  \pg{JTST0003--JTST0003}.

\bibitem[Viotti {\em et~al.\/}(2014)Viotti, Quadrio \& Luchini]{15}
{\sc \au{Viotti, C.}, \au{Quadrio, M.} \& \au{Luchini, P.}} \yr{2014}  \at{Streamwise oscillation of spanwise velocity at the wall of a channel for turbulent drag reduction}.  \jt{Physics of Fluids}  \bvol{26}~(10),  \pg{101504}.

\bibitem[Vreman \& Kuerten(2014)]{DNS2}
{\sc \au{Vreman, A.~W.} \& \au{Kuerten, J. G.~M.}} \yr{2014}  \at{Comparison of direct numerical simulation databases of turbulent channel flow at $re_\tau = 180$}.  \jt{Physics of Fluids}  \bvol{26},  \pg{015102}.

\bibitem[Yamamoto {\em et~al.\/}(2013)Yamamoto, Hasegawa \& Kasagi]{DHT1}
{\sc \au{Yamamoto, A.}, \au{Hasegawa, Y.} \& \au{Kasagi, N.}} \yr{2013}  \at{Optimal control of dissimilar heat and momentum transfer in a fully developed turbulent channel flow}.  \jt{Journal of Fluid Mechanics}  \bvol{733},  \pg{189–220}.

\bibitem[Yuan {\em et~al.\/}(2019)Yuan, Zhang, Cui \& Khoo]{Wenjun_et_al_2019}
{\sc \au{Yuan, W.}, \au{Zhang, M.}, \au{Cui, Y.} \& \au{Khoo, B.~C.}} \yr{2019}  \at{Phase-space dynamics of near-wall streaks in wall-bounded turbulence with spanwise oscillation}.  \jt{Physics of Fluids}  \bvol{31}~(12),  \pg{125113},  \arxiv{arXiv: https://pubs.aip.org/aip/pof/article-pdf/doi/10.1063/1.5130161/13661448/125113\_1\_online.pdf}.

\bibitem[Zhang {\em et~al.\/}(2020)Zhang, Li \& Jiang]{6}
{\sc \au{Zhang, X.}, \au{Li, T.} \& \au{Jiang, N.}} \yr{2020}  \at{Flow control strategies for improving performance of turbomachinery: A review}.  \jt{Progress in Aerospace Sciences}  \bvol{115},  \pg{100595}.

\end{thebibliography}

\end{document}